\documentclass[12pt]{article}
\usepackage{graphicx} 
\usepackage[margin=2.5cm]{geometry}
\usepackage{xcolor}
\usepackage{amsmath}
\usepackage{amssymb}
\usepackage{booktabs}
\DeclareMathOperator*{\argmin}{arg\,min}
\usepackage[sortcites=true, sorting=none]{biblatex}
\usepackage{makecell}
\usepackage{subfigure}
\usepackage[utf8]{inputenc}[]
\title{Parameter-Efficient Fine-Tuning of Machine-Learning Interatomic Potentials for Phonon and Thermal Properties}
\usepackage{authblk}
\author[1]{Jonas Grandel}
\author[1,*]{Philipp Benner}
\author[1,2,*]{Janine George}
\affil[1]{Federal Institute for Materials Research and Testing, Berlin, Germany}
\affil[2]{Institute of Condensed Matter Theory and Solid-State Optics, Friedrich Schiller University Jena, Germany}
\affil[*]{Correspondence should be addressed to: janine.george@bam.de and philipp.benner@bam.de}

\date{March 2026}

\begin{document}

\maketitle

\section{Abstract}
Machine-learning interatomic potentials are widely used as computationally efficient surrogates for density functional theory in atomistic simulations, enabling large-scale, long-time modeling of materials systems. We investigate how different fine-tuning strategies influence the prediction of harmonic phonon band structures, thermal properties, and the potential energy surface along imaginary phonon modes. We achieve substantial accuracy improvements with minimal additional data, with as few as 10 additional training structures already yielding significant gains. In addition to existing approaches, we introduce \emph{Equitrain}, a fine-tuning framework that implements LoRA-based adaptation. Across 53 materials systems, we show that fine-tuned models consistently outperform both the underlying pretrained model and models trained from scratch. Equitrain achieves the best overall performance, and our results demonstrate that fine-tuning enables accurate phonon predictions.

\section{Introduction}
\begin{figure}[ht]
    \centering
    \includegraphics[width=0.7\textwidth]{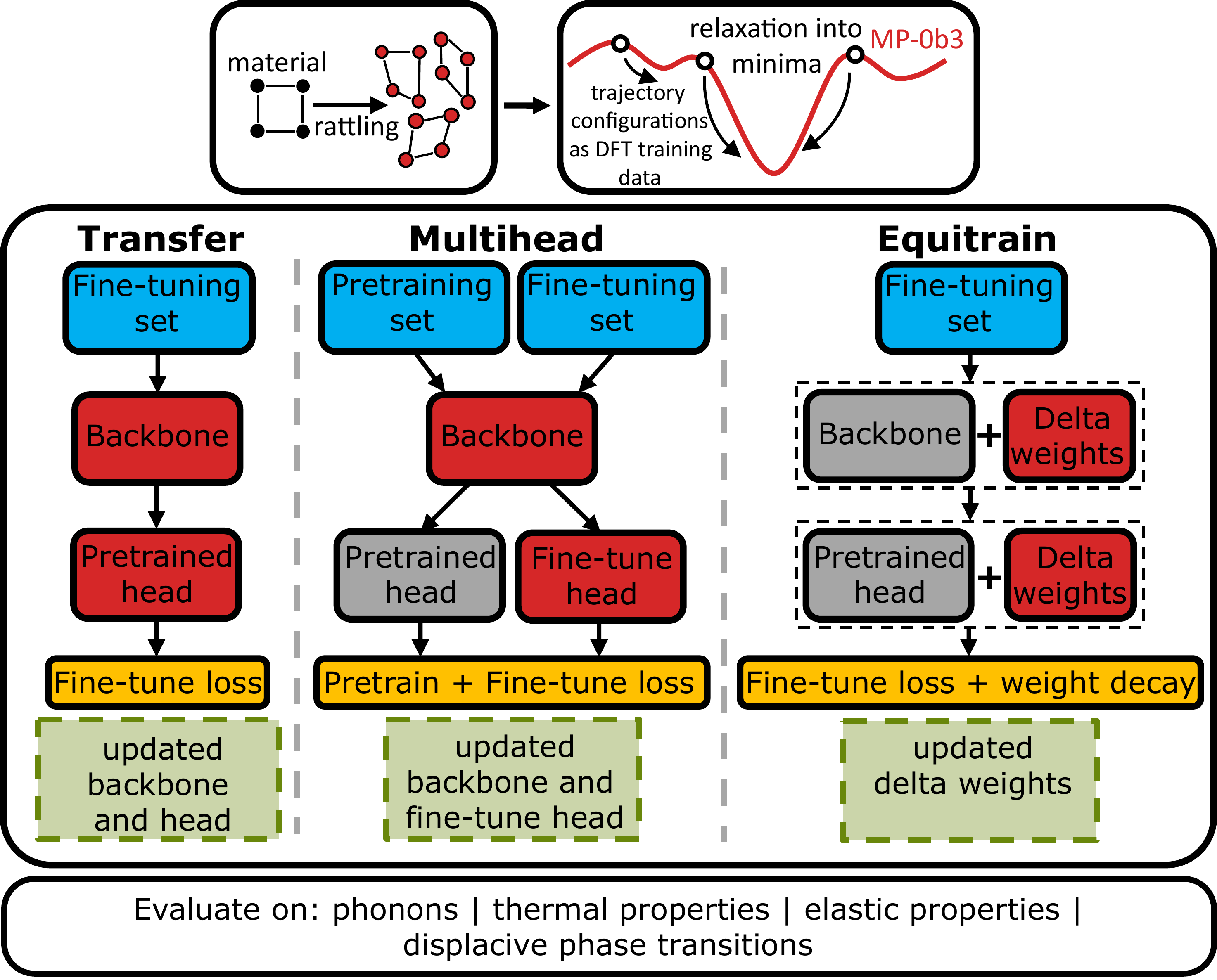}
    \caption{Overview of the concepts and fine-tuning strategies considered in this work. Rattled structures were generated for each material and used to obtain relaxation trajectories with the foundation MP-0b3 model, from which the DFT fine-tuning data set was constructed. With this, different fine-tuning strategies were tested and evaluated in terms of phonons, thermal and elastic properties and phase transition behavior.
    Blue boxes indicate the training data, red boxes highlight the model components updated during fine-tuning, and gray boxes represent components that remain frozen.
    Transfer learning (left) updates both the pre-trained backbone and task-specific head using only the fine-tuning objective. Multihead (center) jointly optimizes a shared backbone using both pretraining and fine-tuning losses, while maintaining separate task heads. Equitrain (right) parametrizes the backbone and output head weights as $\omega = \omega_0 + \Delta\omega$. A weight decay term is applied exclusively to $\Delta\omega$, regularizing deviations from the pre-trained initialization. }
    \label{fig:overview}
\end{figure}
Pre-trained machine-learned interatomic potentials (MLIPs) such as MACE-MP-0b3 \cite{batatia2025foundation} are increasingly used to predict ground-state properties and to perform both molecular dynamics and static calculations. Due to recent advances in model architectures and training strategies, modern MLIPs now achieve accuracies comparable to density functional theory (DFT), making them a promising alternative for large-scale or high-throughput calculations. Previous studies have trained MLIPs explicitly for phonon calculations using large datasets \cite{elena2025machine, lee2025accelerating}. In contrast, we investigate whether accurate phonon predictions can be achieved through fine-tuning with only a small number of additional training structures. When trained on large and diverse collections of materials spanning a wide range of crystal structures and chemical compositions, such pre-trained MLIPs can be viewed as foundation models. In this setting, they provide transferable representations of atomic interactions that generalize across chemical and structural spaces, enabling accurate predictions for materials beyond those explicitly seen during training \cite{batatia2025foundation}.
For certain properties, however, stringent accuracy requirements remain. Phonon calculations, in particular, are highly sensitive to force errors on the order of 1 meV/\AA, making them a challenging benchmark for machine-learning models.
While certain harmonic-phonon properties, such as heat capacity, already show very good agreement with DFT reference data \cite{loew2025universal}, accurate prediction of phonon band structures, especially dynamic stabilities, remains challenging. In particular, the correct prediction of an imaginary mode by an ML model does not necessarily imply that the underlying potential energy surface (PES) along the unstable direction is accurately represented. \\
Figure \ref{fig:overview} provides a schematic overview of the workflow and the key methodological components addressed in this work. 
A variety of fine-tuning strategies \cite{batatia2025foundation, radova2025fine} have been invented and tested to adapt foundation models effectively for systems of interest. 
It has been successfully demonstrated that fine-tuning can reduce systematic errors inherent to foundation models \cite{deng2025systematic}. 
Naive transfer learning, however, carries the risk of catastrophic forgetting \cite{mccloskey1989catastrophic}, as model parameters are directly updated on a new system-specific dataset. As a result, the learned parameters may drift away from those optimized on the much larger, more diverse pretraining dataset, thereby degrading the model's generalization capabilities.
One approach to mitigate this issue is to freeze single model layers during fine-tuning \cite{radova2025fine}. This can be particularly useful for large fine-tuning datasets, accelerating training. 
An alternative is the multi-head fine-tuning approach, in which a subset of the foundation model's training data is used as a replay set alongside the new fine-tuning data \cite{batatia2022mace, batatia2025foundation}. During training, all backbone weights are optimized across both datasets, ensuring parameter updates remain close to the foundation model. Only the output head is trained from scratch. The main drawback of this method is the increasing computational cost as the size of the replay set grows.
An approach that combines fast training and avoids excessive deviation of the model parameters from the foundation model baseline is the LoRA method \cite{hu2022lora} that we implemented in a modified form for this study in a software package called Equitrain \cite{equitrain_website}. We freeze the pre-trained weights and only train additive parameters. Only those and the task head are updated during fine-tuning. A weight decay term is applied exclusively to the additive parameters, regularizing deviations from the pre-trained initialization.
These different strategies correspond to the fine-tuning approaches summarized in \ref{fig:overview}.\\
Against this methodological background, predicting phonon properties represents a particularly relevant and demanding test case.
Phonon calculations are computationally expensive. In the finite displacement approach, as it is implemented in the Phonopy code \cite{phonopy-phono3py-JPCM, phonopy-phono3py-JPSJ}, single atoms in supercells are displaced to calculate all force constants of the system. This requires large supercells to obtain converged results. Other approaches exist for calculating force constants using rattled structures \cite{eriksson2019hiphive}. MLIPs can dramatically reduce the computational cost of such calculations, but their predictive quality for vibrational properties is not guaranteed. By adding only a small number of additional training structures, fine-tuning offers the prospect of retaining most of the computational savings while substantially improving accuracy. \\
Since we base our phonon calculations on single-atom displaced supercells, the most naive training dataset would consist of such cells. However, since the MACE architecture relies on (semi-)local atomic environments, single-atom-displaced supercells as training data provide limited direct information, as almost all atoms are in their equilibrium positions. 
To capture the energetics relevant to phonon calculations, we focus on accurately fitting the PES near the energy minimum. As illustrated in Figure \ref{fig:overview}, we generate 
scaled and deformed rattled structures, relax them using the foundation MACE-MP-0b3 model (further referred to as MP-03b), and select configurations uniformly spaced in energy to sample the energy landscape around equilibrium efficiently.
Moreover, in contrast to the typical approach to fine-tuning a single model across multiple materials \cite{elena2025machine, lee2025accelerating}, we train an individual model for each material.
As part of this procedure, we also examine how the size of the training structures influences phonon predictions.\\
While phonon mean absolute errors or extremal frequencies are commonly used as global indicators of phonon prediction quality, such metrics provide only a statistical view and do not capture how phonon errors propagate into specific physical properties.
In this work, we systematically evaluate different fine-tuning strategies for phonon predictions across 53 materials, with particular emphasis on the LoRA-based framework Equitrain introduced here.
Beyond phonon frequencies, we assess the performance of different fine-tuning strategies for thermal properties, elastic properties, and the equations of state. Additionally, we also investigate the extent of catastrophic forgetting and analyze imaginary phonon modes as a stability criterion, examining how well the PES is represented along unstable directions.

\section{Results and discussion}
\subsection{Fine-tuning strategies}
Fine-tuning is required to adapt pre-trained foundation MLIPs to new materials systems or target properties using limited additional data. In this study, we compare several fine-tuning strategies for adapting such models, including transfer learning, multi-head fine-tuning, and LoRA-based adaptation. For transfer learning and multi-head fine-tuning, we use the MACE package \cite{batatia2022mace, batatia2025design}. Multi-head fine-tuning mitigates catastrophic forgetting by replaying data from the original training set during fine-tuning; however, it incurs a substantial computational overhead. In addition, we implemented our own LoRA-based fine-tuning strategy in \emph{Equitrain} \cite{equitrain_website}, a software package for efficient training and fine-tuning of MLIPs. LoRA allows efficient adaptation of machine learning models by training additive weight updates while keeping the base model fixed. In this study, we use full-rank LoRA as outlined and justified in the Methods section.

\subsection{Data efficiency of fine-tuning strategies}
\begin{figure}[ht]
    \centering
    \subfigure[]{\includegraphics[width=0.49\textwidth]{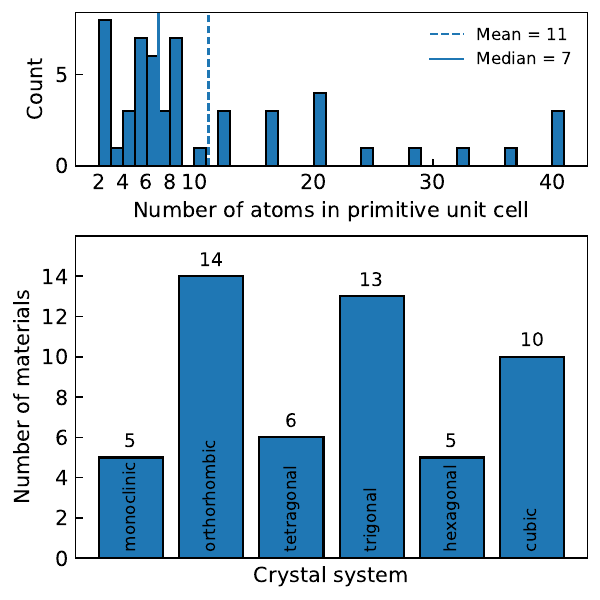}}
    \subfigure[]{\includegraphics[width=0.49\textwidth]{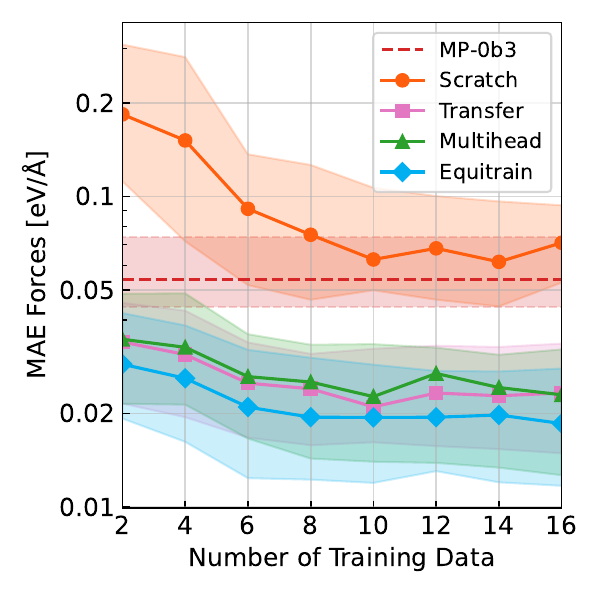}}
    \caption{(a) Distribution of crystal systems and the number of atoms per primitive unit cell in the dataset, which contains a total of 53 materials. (b) Median MAE forces errors across all 53 material models within the IQR, as shown in the shaded area in the background. Note that the y-axis is logarithmic, and lines are drawn between data points for improved readability. The performance of the foundation model, MP-0b3, on the validation data is represented by a dashed horizontal line for comparison.}
    \label{fig:forces_supercell_convergence}
\end{figure}

\begin{figure}[h!]
    \centering
    \includegraphics[width=0.5\linewidth]{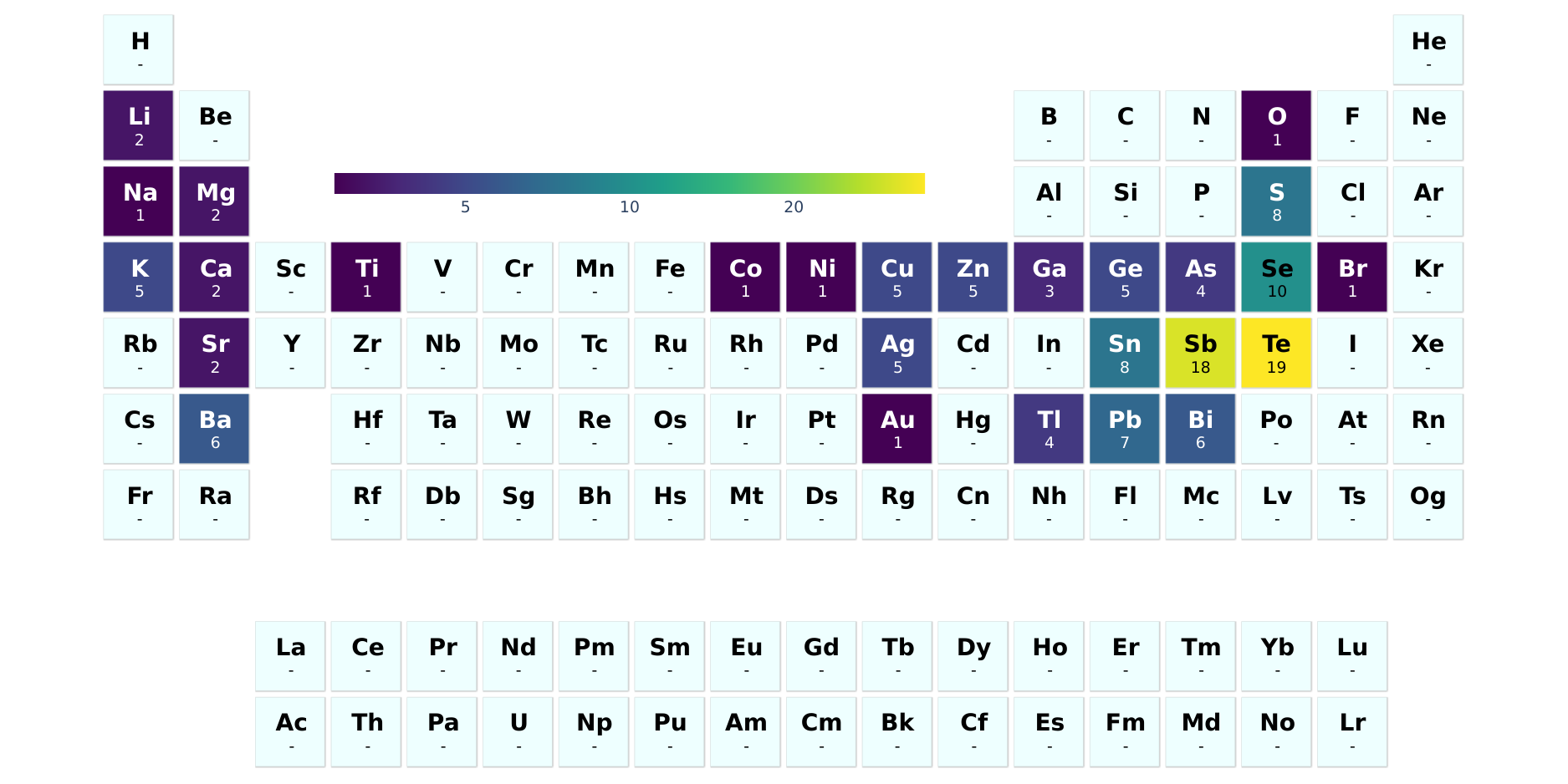}
    \caption{Heatmap of elements contained in the material dataset. The heatmap was created with \cite{riebesell_pymatviz_2022}.}
    \label{fig:SI_element_heatmap}
\end{figure}

For a rigorous evaluation, we calculate phonon band structures for 53 materials.
Figure \ref{fig:forces_supercell_convergence}a summarizes the distribution of crystal systems and the number of atoms per unit cell in the dataset. The chosen materials cover all crystal systems except the triclinic one and include not only binary or ternary compounds but also structurally more complex systems with up to 40 atoms per unit cell. The dataset is dominated by phase-change materials and chalcogenide-based semiconductors reflecting their relevance for thermoelectric applications, in which phonon properties play a central role \cite{pandey2022updates}. Most materials contain tellurium, antimony, or selenium, and the element heatmap is shown in SI \ref{fig:SI_element_heatmap}. All materials were selected from the Materials Project database \cite{jain2013commentary, horton2025accelerated}.\\
In a first step, we determine how much additional, system‑specific training data is required for fine-tuning. Because our main interest lies in harmonic phonon calculations, we focus on fine‑tuning the models using configurations sampled in the vicinity of the equilibrium structure so that a harmonic potential can be approximated well by the MLIP. 
The training data comprise rattled structures and selected configurations sampled along relaxation trajectories obtained using the MP-0b3 foundation model. Detailed information on the data generation protocol is provided in the Methods section.
To quantify data efficiency, we systematically increase the number of training configurations and fine-tune an individual model per material and strategy. We examine the energy, force, and stress prediction errors for the individual validation sets of each model across different fine-tuning strategies. 
Because phonon properties are particularly sensitive to force accuracy, we focus here on force mean absolute errors (MAE). The corresponding energy and stress results are consistent with the same overall trend across all training structure sizes and are listed in the supplementary information.
Figure \ref{fig:forces_supercell_convergence} (b) compares the force MAE achieved by different fine-tuning strategies when trained on large-supercell datasets.
For transfer learning, multi-head fine-tuning, and Equitrain, the median force MAE decreases and converges fast as the training set grows. The shaded region shows the interquartile range (IQR), highlighting the variation in fine-tuning performance across materials.
Relative to the MP-0b3 foundation model, all fine-tuning approaches based on pre-trained weights achieve a clear improvement in force MAE. Models trained from scratch perform worst overall, highlighting the advantage of pre-trained models when training data are limited. 
Within all approaches, \emph{Equitrain} demonstrates the best overall performance, followed by transfer learning and multi-head fine-tuning. The comparatively slightly inferior performance of the latter likely results from the additional, non-system-specific replay dataset during fine-tuning.
For the frozen-layer approach \cite{radova2025fine}, we observe only small changes in performance across different numbers of frozen layers. In the limit of zero frozen layers, this leads ultimately to the transfer learning approach. Since the training data sets are small, there is no significant advantage in the observed training speed. We therefore do not consider frozen-layer fine-tuning further in this work (see also SI \ref{SI:freezing-layer}). 
Based on this analysis, we restrict the following evaluation to transfer learning, multi-head fine-tuning, \emph{Equitrain}, and from-scratch models trained on 10 configurations per material and compare against the foundation MP-0b3 model.

\subsection{Effect of training structure size on phonon accuracy}
We next assess how the size of the training structures affects phonon accuracy.
For this, we test primitive unit cells, small cubic supercells with a maximal edge length of 10 \AA{}, and large cubic supercells with a maximal edge length of 15 \AA {}.
To evaluate the quality of the phonon band structure calculations within the different fine-tuning strategies, we measure the MAE between the DFT-calculated and the ML-calculated phonon frequencies in the whole Brillouin zone. The results are shown in Table \ref{table:phonon_mae}.
\begin{table}[ht]
    \centering
    \caption{Median phonon MAE and IQR in THz of all 53 material models within different fine-tuning strategies and different training structure sizes. \\}
    \begin{tabular}{c |c c c c} 
        \hline
                        & Scratch & Transfer & Multihead  & Equitrain \\ \hline
        primitive       & 0.39 (0.50) & 0.12 (0.07) & 0.18 (0.16)  & 0.12 (0.12)\\
        small supercell & 0.25 (0.39) & 0.09 (0.06) & 0.12 (0.06)  & 0.07 (0.06)\\
        large supercell & 0.21 (0.14) & 0.06 (0.05) & 0.07 (0.06)  & \textbf{0.05 (0.05)}\\ \hline
    \end{tabular}
    \label{table:phonon_mae}
\end{table}
In the case of the primitive and small supercell from-scratch models, phonons could not be calculated for two materials because the structures relaxed into nonphysical configurations.
All fine-tuning techniques achieve significant improvements in predicting phonon band structures. Two main trends can be identified in the data.
First, we observe a decrease in overall MAE from small to large training structures. This behavior is expected because the number of atoms, and therefore the number of diverse local environments that the model encounters during training, increases in larger structures. Additionally, larger supercells capture long-range force information, which is necessary for accurate phonon calculations. \\
Secondly,  \emph{Equitrain} is the most successful strategy and achieves the lowest median phonon MAE. For comparison, the foundation MP-0b3 model shows a median MAE (IQR) of 0.27 THz (0.18 THz), which is more than five times higher than the Equitrain approach trained with large supercells. Additional comparisons with foundation models trained on larger data sets can be found in SI \ref{SI:diff_foud_models}.\\
Given the significant performance advantage of large-supercell training, all subsequent comparisons are limited to models trained on large supercells. We start by comparing the maximum and mean phonon frequencies, and the density of states reported in Table \ref{table:phonon_metric}.
\begin{table}[ht]
    \centering
    \caption{Median error and IQR of maximal phonon frequencies, ($\omega_{\text{max}}$), mean phonon frequencies ($\omega_{\text{mean}}$) and density of states ($\Delta$DOS) for large-supercell training data.\\}
    \begin{tabular}{c |c c c c c} 
        \hline
        &         MP-0b3 & Scratch &  Multihead & Transfer & Equitrain \\ \hline
        $\Delta \omega_{\text{max}}\,[\%]$  & -7.0 (9.0) &  -7.9 (14.4) & -3.1 (4.5)  & -1.6 (4.2) & \textbf{-0.9 (3.8)} \\
        $\Delta \omega_{\text{mean}}\,[\%]$ & \textbf{2.8 (1.9)} &  3.2 (2.0) & 3.0 (1.8)  & 2.9 (1.7) & 3.0 (1.7) \\
        $\Delta$DOS [\%]                    & 50.8 (33.1) & 55.7 (22.8) & 19.6 (15.6)  & 20.3 (13.5) & \textbf{14.2 (14.5)} \\ \hline
    \end{tabular}
    \label{table:phonon_metric}
\end{table}
Negative signs in the table indicate that the ML approaches are underestimating the reference results.  
Overall, the fine-tuning approaches show smaller or comparable errors to the foundation MP-0b3 model, while the from-scratch models perform worst, as expected, across all three quantities.
For the prediction of $\omega_{\text{max}}$, the \emph{Equitrain} approach performs almost twice as well as the transfer learning approach and three times as well as the multi-head strategy. 
A main reason for the poor MAE of the foundation MP-0b3 model is a systematic underprediction of maximum phonon frequencies due to systematic softening in foundational MLIPs \cite{deng2025systematic}. Forces are underestimated, resulting in a smaller $\omega_{max}$ and thus a poorer MAE. By fine-tuning, the softening effect is compensated to varying degrees, depending on the strategy. A similar underestimation of phonon frequencies was observed in our earlier study, where phonon properties were computed directly using the MP-0b3 foundation model \cite{batatia2025foundation}. Parts of the benchmark data overlap with the dataset used in that study.
For $\omega_{\text{mean}}$, all models perform equally well. This suggests that, to calculate properties that depend only on averaged frequencies, the foundation model is already sufficient.
Compared to the foundation model, all fine-tuned models show improved predictions of the phonon density of states (DOS), with the Equitrain approach achieving the best performance.

\subsection{Evaluation of phonon-dependent thermodynamic and elastic properties}
\begin{figure}[ht]
    \centering
    \includegraphics[width=1\linewidth]{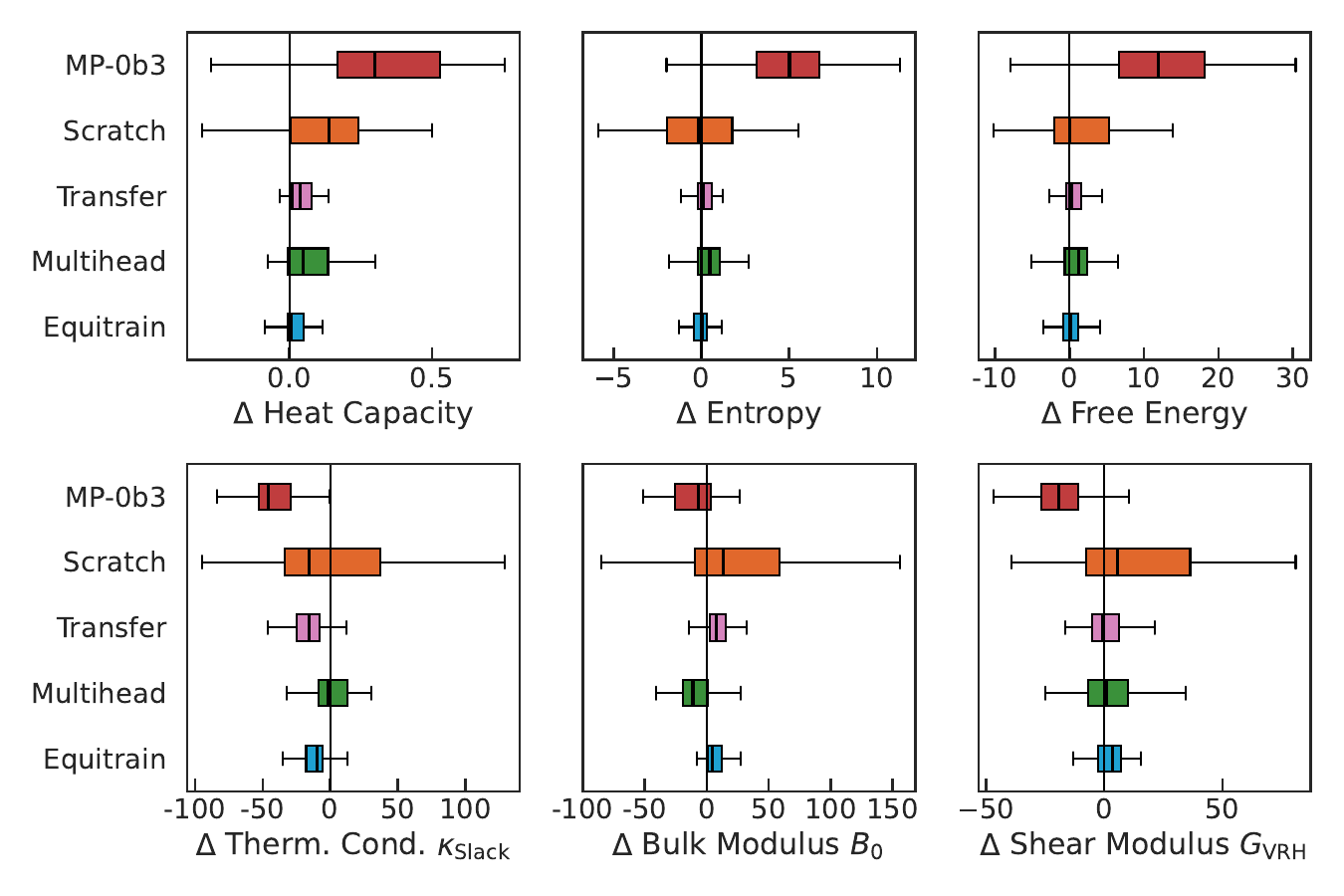}
    \caption{Deviation [\%] of thermal and elastic properties from the DFT reference results. For better readability, outliers are not shown but are included in the SI \ref{SI:phys_prop_with_out}.}
    \label{fig:thermal_prop}
\end{figure}

To assess how the improved phonon descriptions translate into physically relevant quantities, we extend the evaluation to the phonon-dependent properties, namely heat capacity at constant volume, vibrational part of the entropy and Helmholtz free energy. Furthermore, we calculate the elastic properties, such as the shear modulus and bulk modulus, and derive the Slack thermal conductivity \cite{slack1973nonmetallic} from them. The temperature-dependent properties are calculated at 300 K. \\
We classify the Slack thermal conductivity as an elastic property here, since it depends on the average Grüneisen parameter, which we derive from the longitudinal and transversal sound velocities $v_L, v_T$ \cite{jia2017lattice} obtained from elastic constants \cite{jia2017lattice, ganose2025atomate2} (see also SI \ref{SI:slack_thermal_cond}).
In this sense, thermal conductivity, shear modulus, and bulk modulus provide a good test of how well the MACE models generalize, as they are sensitive to the quality of the energy predictions under small deformations such as compression or shear. \\
Figure \ref{fig:thermal_prop} summarizes the deviations of ML-predicted and DFT reference values. 
Across all phonon-dependent properties, fine-tuned models consistently outperform both the foundation MP-0b3 model and models trained from scratch.
All fine-tuning strategies exhibit narrow error distributions with median deviations close to zero, indicating only small systematic bias.  
Among them, \emph{Equitrain} performs best overall, with the vast majority of materials falling within a $\pm 5\%$ error range across all phonon-dependent properties. Transfer learning achieves comparable accuracy, whereas the multi-head approach shows slightly larger dispersions. 
Since an additional replay set is used for multi-head fine-tuning to prevent catastrophic forgetting, we conclude that adding this data during fine-tuning is at least unnecessary and slightly counterproductive for predicting phonon-dependent properties.
Interestingly, the from-scratch models show improvements over the foundation MP-0b3 model. This is in contrast to the earlier-found errors on the phonon band structure predictions themselves. It shows that band-structure errors do not directly translate into errors in thermodynamic properties.\\
Among the elastic properties, all fine-tuning models show improvements over the foundation MP-0b3 model.
In contrast to the phonon-dependent properties, the poor performance of the from-scratch models indicates that elastic responses cannot be learned reliably from limited training data.
Transfer learning and \emph{Equitrain} tend to underestimate thermal conductivity, corresponding to underestimated sound velocities and overestimated Grüneisen parameters. \\
For the bulk and shear moduli, both methods yield comparable improvements, while the multi-head approach exhibits a slightly broader error distribution. \\
Across all thermodynamic and elastic properties, the foundation MP-0b3 model exhibits substantial systematic deviations. It systematically underestimates the thermal conductivity by around $-50$\% on median, while it shows a wide error dispersion. From-scratch models show slightly improved performance on thermal properties but fail to predict elastic properties. They suffer from either systematic offsets or comparably large dispersions, making them much less reliable than any of the fine-tuning strategies.
Overall, these results highlight that fine-tuning is the most reliable route to accurate thermodynamic and elastic properties, with \emph{Equitrain} and transfer learning providing the most robust and consistent performance.

\subsection{Imaginary mode prediction and phase transition}
\label{results_imaginary_mode_pathways}
An even more stringent test of model fidelity is the prediction of dynamical instabilities. Phonons probe the curvature of the PES beyond the harmonic minimum and are directly linked to structural phase transitions. Identifying imaginary modes is particularly challenging because it requires a faithful description of shallow instabilities and anharmonic effects, and is therefore challenging not only for ML models but also for DFT calculations themselves.
Computational inaccuracies can introduce non-physical instabilities \cite{pallikara2022physical} and consequently, a direct comparison between DFT and ML predictions based solely on the existence of imaginary modes is delicate. \\
We first evaluate dynamical stability only at commensurate $k$-points, with a stringent cutoff frequency of $-10^{-4}$ THz.
\begin{table}[ht]
    \centering
    \caption{Confusion matrix of stable or unstable structures in the dataset indicated by imaginary modes at commensurate $k$-points. According to the DFT data, 44 structures are stable, while 9 are unstable. Precision and recall are given with respect to imaginary modes. \\}
    \begin{tabular}{c | c c c c c} 
        \hline
        &           MP-0b3 & Scratch & Transfer  & Multihead & Equitrain \\ \hline 
        true stable         & 42 & 43 & 43 &42 & 44\\
        false stable        &  2 & 1 & 1 & 2  &  0\\
        false unstable      &  2 & 6 & 3 & 1  &  1\\ 
        true unstable       &  7 & 3 & 6 & 8  &  8\\ \hline
        neg. Precision      &  78\% & 75\%  & 86\%  & 80\%  & 100\% \\
        neg. Recall         &  78\% & 33\%  & 67\%  & 89\%  & 89\% \\
        
    \end{tabular}
    \label{table:imaginary_modes}
\end{table}

\begin{figure}[h!]
    \subfigure[]{\includegraphics[width=1\textwidth]{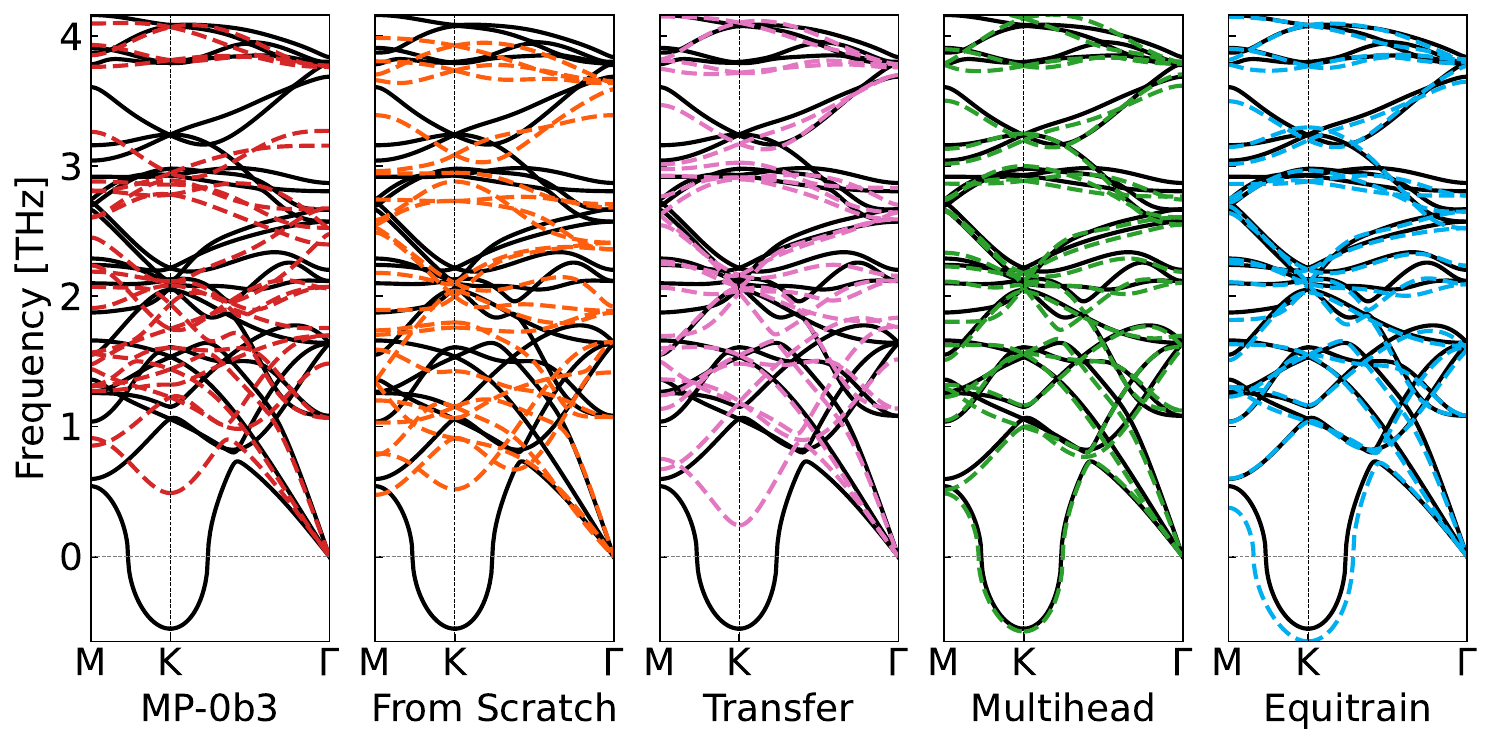}}
    \subfigure[]{\includegraphics[width=0.4\textwidth]{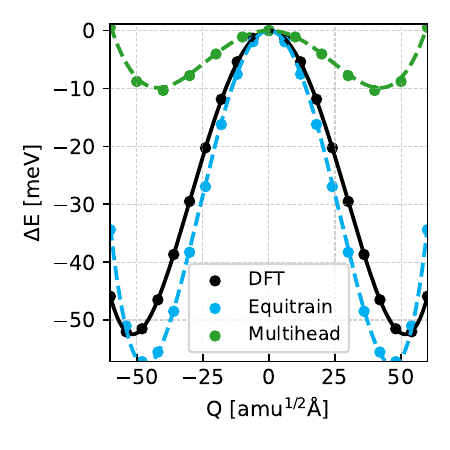}}
    \subfigure[]{\includegraphics[width=0.6\textwidth]{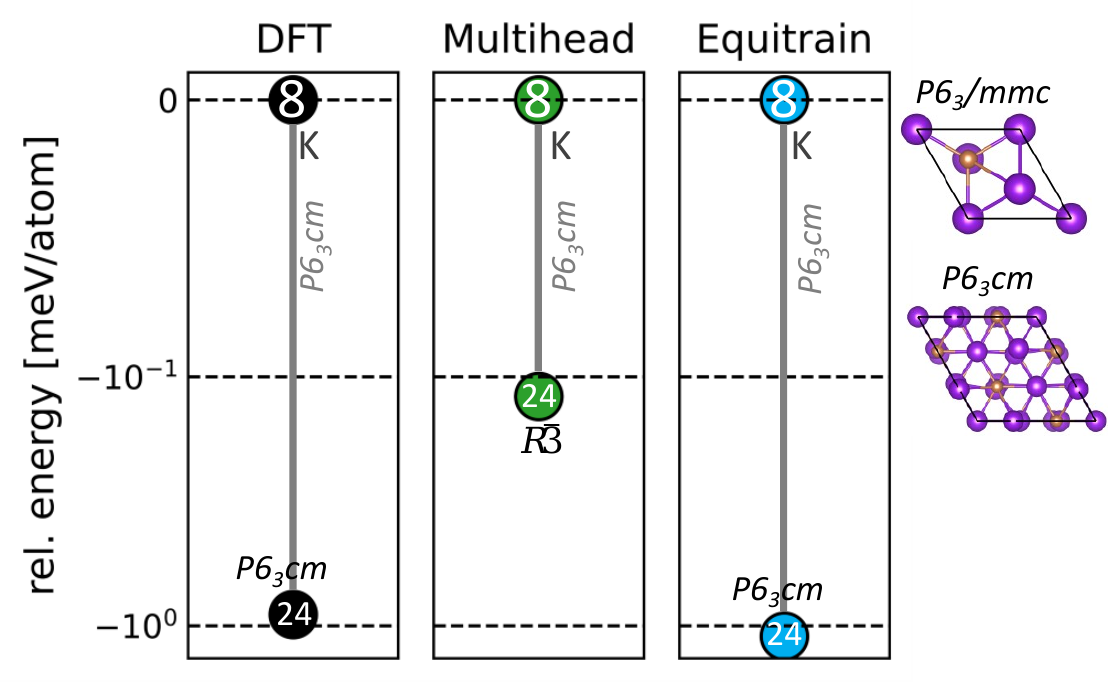}}
\caption{(a) Section of the phonon band structure of K$_3$Sb calculated with the ML models and compared to DFT (black). (b) Comparison of the anharmonic double-well potential along the $K$ mode. (c) Phase transition pathway of the models showing imaginary modes.}
\label{fig:mp-14017}
\end{figure}

\begin{figure}
    \subfigure[]{\includegraphics[width=0.7\textwidth]{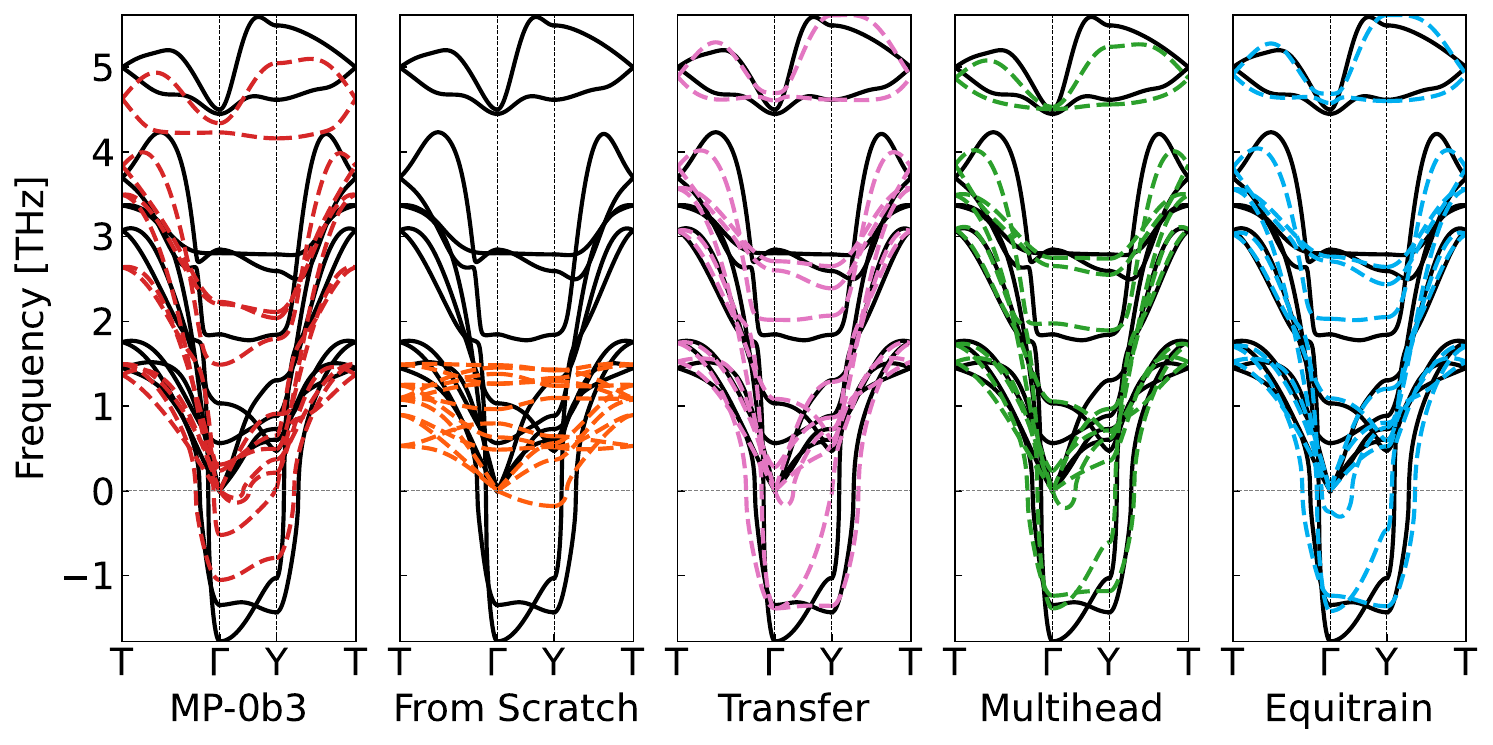}}
    \subfigure[]{\includegraphics[width=0.28\textwidth]{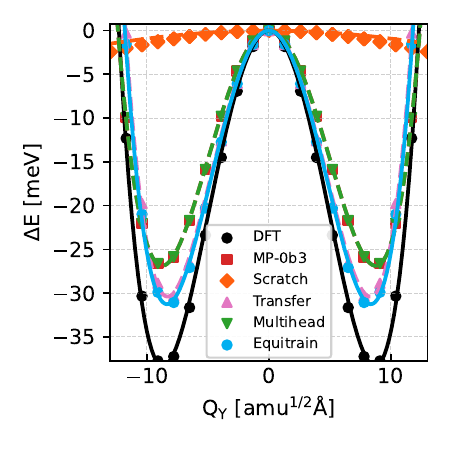}} 
    \subfigure[]{\includegraphics[width=1\textwidth]{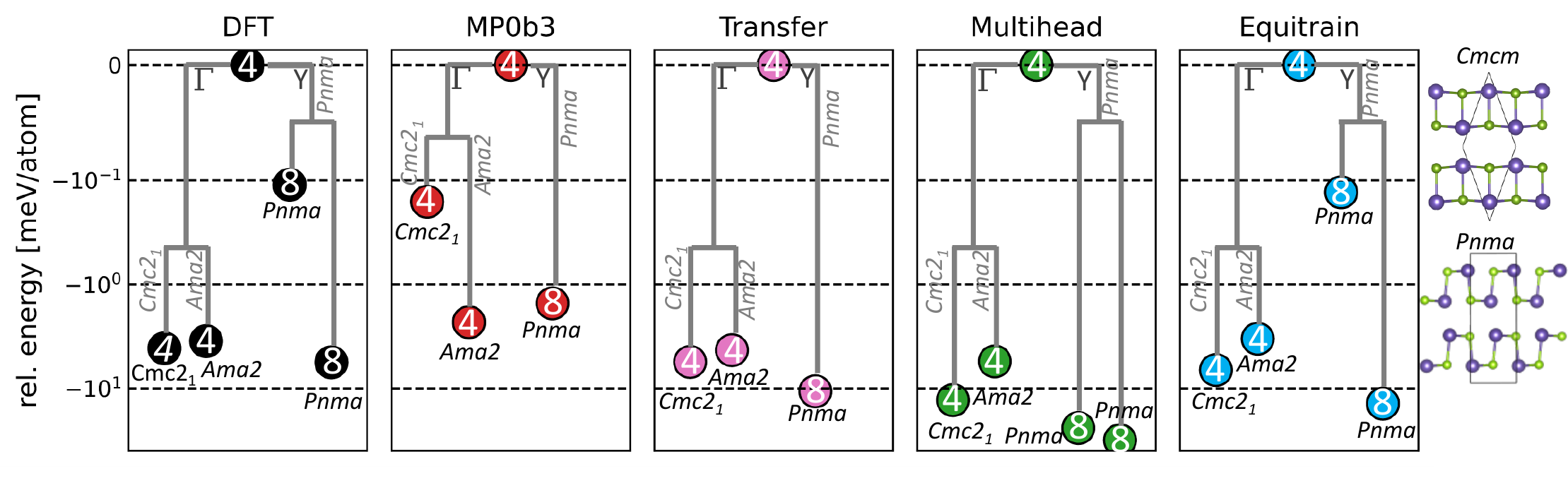}}
\caption{(a) A section of the phonon band structure calculation of \textit{Cmcm} SnSe with the different ML models compared to the DFT benchmark (black). (b) Comparison of the anharmonic double-well PES of the lowest $Y$-mode. (c) Phase transition pathways along the $\Gamma$ and $Y$ imaginary modes.}
\label{fig:mp-2168}
\end{figure}

The DFT dataset contains nine unstable and 44 stable structures. The performance of the ML models in identifying unstable structures is shown in the confusion matrix in Table \ref{table:imaginary_modes}. 
Among all ML approaches, the Equitrain strategy shows the best agreement with the DFT benchmark. It achieves a negative precision of 100\%, meaning that no stable material is misclassified as unstable, and a negative recall of 89\%, with only a single unstable material incorrectly predicted as stable. The multi-head fine-tuning models perform second best, incorrectly predicting two materials as stable and one as unstable. The foundation MP-0b3 model slightly outperforms the transfer learning models, while the from-scratch models show substantial difficulty in reproducing the correct stability classification.
We conclude that additional training data, present either in the fine-tuned models due to the foundation model itself or during multi-head fine-tuning using replay data, has a beneficial effect on predicting instabilities. From-scratch models lack this additional data and therefore often fail to find alternative energy minima close to the relaxed structure.
Transfer models tend to exhibit critical forgetting and, therefore, perform slightly worse than the other two fine-tuning approaches.
\\
However, detecting an imaginary mode is not sufficient on its own. The most stringent question is whether the model also reproduces the correct anharmonic pathway and the final displacive phase.
To address this, atoms are collectively displaced along selected imaginary modes, and the potential energy $U$ is evaluated at different fixed mode amplitudes along the deformation, following the procedure proposed in \cite{togo2013evolution}. In this way, new energy minima and polymorphs can be identified.
For the nine unstable materials, imaginary modes at high-symmetry points as well as band-structure minima are selected, and commensurate supercells are constructed only when their size remains below 1000 atoms. Because of the size restriction, DFT reference calculations are feasible for only 26 out of 41 (63\%) considered $k$-points.
\\
K$_3$Sb provides a representative example in which the relevant instability must be identified and followed to the correct polymorph.
Figure \ref{fig:mp-14017}(a) shows a section of the phonon band structure calculated with DFT and the ML models. The DFT reference exhibits a prominent imaginary mode at $K$. MP-0b3, from-scratch and transfer models fail to reproduce this instability, whereas the multi-head and Equitrain models capture it correctly. By following the mode, we obtain the potential energy surface $U(Q)$ along the normal coordinate $Q$ as shown in Figure \ref{fig:mp-14017}(b).
Both multi-head and Equitrain reproduce the qualitative shape of the double-well potential, but differ noticeably in the depth and position of the minima.
In both cases, the minima are shifted inward relative to DFT, while Equitrain overestimates the energy difference and the multi-head model significantly underestimates it. The absence of an imaginary mode at $K$ in the phonon band structures of the other models corresponds to a harmonic potential, which is inconsistent with the DFT reference. 
The crucial test, however, is whether ML models can correctly predict phase transition pathways. Relaxing the configurations in the energy minima of the double-well potentials leads to a new phase. For K$_3$Sb, the phase transition pathway along $K$ is shown in Figure \ref{fig:mp-14017}(c). Only the Equitrain model recovers the relaxed structure in the same space group symmetry of $P6_3cm$, while the multi-head model fails and relaxes to a different phase with space group symmetry $R\bar{3}$.
\\
As a second example, Figure \ref{fig:mp-2168} shows a section of the phonon band structure of SnSe together with the corresponding phase-transition pathways. 
The high-temperature $Cmcm$ phase of SnSe is known to undergo a phonon-induced phase transition, leading to the low-temperature $Pnma$ phase \cite{skelton2016anharmonicity}.
Here, all fine-tuned models are in good agreement with the DFT phonon benchmark. The from-scratch approach fails completely, and the foundation MP-0b3 model shows the characteristic softening in the higher phonon frequencies. All models show a small additional imaginary mode between $\Gamma$ and $Y$ which does not appear in the DFT calculation.\\
The potential energy $U$ of the lowest $Y$ mode is shown in Figure \ref{fig:mp-2168}(b). All ML models fail to capture the depth of the anharmonic potential, and the minima are shifted inwards. The from-scratch model does not reproduce the shape of $U$ at all. For the DFT reference, we find that, at the $\Gamma$ point, the unstable $Cmcm$ phase relaxes into different symmetry groups $Cmc2_1$ and $Ama2$ depending on the imaginary mode considered.
Figure \ref{fig:mp-2168}(c) shows these phase transition pathways along the different modes at $\Gamma$ and $Y$. The phase transition to the low-temperature phase $Pnma$ is dominated by the smallest imaginary mode at $Y$ \cite{aseginolaza2019phonon}, which is reproduced by all models. Equitrain shows the best agreement in energy of the individual phases. 
\\
Finally, we assess the models more generally by quantifying how many displacive phase transitions can be reproduced across the nine preselected unstable materials. For each material, the set of final relaxed phases predicted with the ML models is compared to the corresponding DFT benchmark set. 
Prior to the evaluation, all phases equivalent to the initial phase are removed, because they represent no new transition. Cases in which DFT exhibits imaginary modes due to interpolation errors within a harmonic potential, while the ML model predicts no instability and therefore provides the physically correct description of the potential energy surface, are excluded from the analysis.
A phase predicted by the ML model is counted as a true positive if it matches a DFT benchmark phase in both space group and number of atoms in the primitive unit cell. Additionally, phases predicted by the ML model but absent in the DFT benchmark are classified as false positives, whereas DFT phases not recovered by the ML model are counted as false negatives.
Further details are provided in SI \ref{SI:phase-transition-pathways}.\\
The resulting precision, recall, and F1 score for the fine-tuning strategies over the set of the nine non-stable structures are presented in Figure \ref{tab:rec_prec_new_phases}. 
\begin{table}[ht]
    \centering
    \caption{Recall, precision, and F1 scores of new phase predictions for different fine-tuning approaches. Precision and recall are calculated separately for each material and then averaged based on the number of materials evaluated.\\}
    \begin{tabular}{c | c c c c c} 
        \hline
        &           MP-0b3 & Scratch & Transfer  & Multihead & Equitrain \\ \hline
        Precision      &  60\%          & 33\%  & 54\%  & 38\%  & $\mathbf{68\%}$ \\
        Recall         &  $\mathbf{66\%}$ & 33\%  & 58\%  & 50\%  & 65\% \\ \hline
        F1             & 0.63 & 0.33 & 0.56 & 0.43 & $\mathbf{0.66}$ \\
    \end{tabular}
    \label{tab:rec_prec_new_phases}
\end{table}
Among the fine-tuned approaches, Equitrain yields the best overall performance, with a precision of 68\%, a recall of 65\%, and an F1 score of 0.66.
Its performance is therefore slightly better than that of the foundation model MP-0b3, which achieves a higher recall of 66\% but a lower precision of 60\%, indicating that it reproduces slightly more benchmark phases at the expense of more false-positive predictions.
In contrast, the other fine-tuning strategies can not compete with these results.
These results suggest that Equitrain best preserves the foundation model's ability to generalize to unseen phases while still benefiting from system-specific fine-tuning. Transfer learning appears to reduce generalization capability, likely due to catastrophic forgetting. Surprisingly, the multi-head approach performs even worse, although the additional replay dataset during fine-tuning is intended to mitigate forgetting of the PES beyond the fine-tuning structures. A possible explanation is that the replay set of 100 structures is too small. However, increasing the size would also increase the computational costs of fine-tuning. 
 
\subsection{Computational cost}
\begin{figure}[ht]
    \centering
    \includegraphics[width=0.6\textwidth]{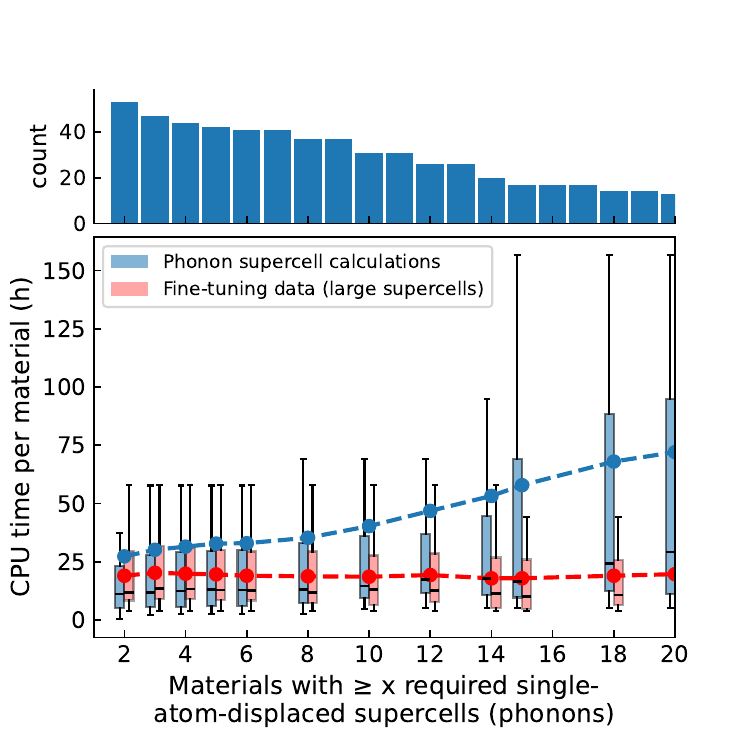} 
    \caption{Average computational cost per material as a function of the minimum required number of single-atom displaced supercells for phonon calculations. The x-axis shows cumulative subsets of materials that require at least a given number of displaced supercells. The blue boxplots correspond to DFT phonon supercell calculations, while the red boxplots indicate the computational cost of generating the large supercell fine-tuning data. The blue and red dashed lines show the mean values. The histogram shows the number of materials counted in each batch.}
    \label{fig:computation_time}
\end{figure}
The computational cost is dominated by the generation of DFT single-atom-displaced supercells required for the phonon calculation, averaging 27.3 h per material (computed using three nodes with 48 CPU cores each). In comparison, generating 10 large supercell training structures for fine-tuning requires 18.8 h per material on average, resulting in an overall reduction in computational time of 32\%. 
This difference reflects the larger size of the single-atom displaced phonon supercells, which have edge lengths above 20 \AA, whereas the large training supercells are typically limited to 15 \AA.
The computational cost of phonon calculations depends strongly on the symmetry of the system, which determines the number of required single-atom-displaced supercells. Figure \ref{fig:computation_time} shows that the benefit of fine-tuning increases for materials requiring at least a given number of single-atom-displaced supercells. For materials requiring 10 or more single-atom-displaced supercells, the total computational time is reduced by 54\% on average, increasing to 92\% for systems requiring 20 or more displaced supercells.
The blue boxplots further show that the spread in CPU time grows with increasing structural complexity. The median runtime for generating single-atom-displaced supercells and fine-tuning data begins to diverge noticeably for materials requiring 12 or more single-atom-displaced supercells. The strongly elevated mean values indicate the presence of outliers that require substantially longer phonon calculations. It is precisely for these cases that the fine-tuning approaches become particularly advantageous. \\
All model training procedures are significantly cheaper, ranging from 1.2 min in scratch and transfer fine-tuning to 7.3 min per material for Equitrain fine-tuning and 8.3 min per material for multi-head fine-tuning with a set of 100 replay structures on a NVIDIA A100 GPU. 

\subsection{Summary}
We systematically evaluated different fine-tuning strategies based on the MACE-MP-0b3 foundation model across phonons, thermal and elastic properties, and phase-transition behavior. Our results show that as few as 10 additional fine-tuning structures are sufficient to substantially improve the predictive performance of the MLIP for a given material. All investigated fine-tuning strategies, i.e., transfer learning, multi-head fine-tuning, and the LoRA-based Equitrain approach, consistently enhance the accuracy of phonon band structures and derived thermal properties compared to both the foundation model and from-scratch training. For elastic properties, a similar trend is observed.
With respect to imaginary phonon modes, Equitrain and multi-head fine-tuning improve instability prediction across the dataset, whereas transfer learning shows reduced performance, most likely due to catastrophic forgetting. However, a more detailed analysis reveals that pure imaginary-mode predictions are not a sufficient indicator of accurate PES prediction. By explicitly evaluating displacive phase transitions, we find that most fine-tuning strategies struggle to generalize across the PES and perform worse than the foundation model. In contrast, Equitrain outperforms the foundation model, indicating that it is the only approach that mitigates catastrophic forgetting and provides the most robust performance across all evaluated properties.


\section{Methods}
\subsection{Reference calculations (DFT)}
\textbf{Electronic structure and relaxations:} All DFT calculations were performed using VASP \cite{kresse1993ab,kresse1996efficient,kresse1996efficiency2} with the PBE exchange–correlation functional \cite{perdew1997generalized} and with the PAW framework\cite{blochl_projector_1994,kresse_ultrasoft_1999}. Unless stated otherwise, an energy convergence criterion of $10^{-8}$ eV, a plane-wave cutoff of 520 eV, and a $k$-point spacing of 0.1 \AA$^{-1}$ were used. Structural relaxations employed a force threshold of 0.5 meV/\AA. \\
\textbf{Phonons and elastic properties:} Harmonic phonons were computed using \texttt{Phonopy} \cite{phonopy-phono3py-JPCM, phonopy-phono3py-JPSJ} with the PhononMaker workflow in \texttt{atomate2} \cite{ganose2025atomate2}. Cubic supercells with a minimum side length of 20 \AA{} were used, sampled at the $\Gamma$ point only. Force constants were extracted using \texttt{symfc} \cite{seko2024projector}. Elastic constants were calculated using the ElasticMaker workflow in \texttt{atomate2} \cite{ganose2025atomate2} with a force threshold of 1 meV/\AA{} and a fixed symmetry during the initial structure relaxation. \\
\textbf{Equation of state:} Equilibrium volumes were scaled between $\pm5\%$ in steps of 1\%, followed by atomic relaxation at fixed volume. Energy–volume data were fitted to the Birch–Murnaghan equation of state \cite{birch1947finite}.

\subsection{Generation of fine-tuning data}
Datasets of primitive unit cells, small and large supercells for fine-tuning were generated using a systematic rattling protocol. Small supercells were constructed as cubic cells with sizes ranging between (5 \AA)$^3$ and (10 \AA)$^3$, while the large cubic supercells ranged from (10 \AA)$^3$ to (15 \AA)$^3$. For each case, five rattled configurations with a mode amplitude of $\sim$0.1 \AA{} were created by using the Monte Carlo rattling procedure implemented in the \texttt{Hiphive} code \cite{eriksson2019hiphive}. The lattice of each rattled structure was further perturbed by randomly scaling the angles and vectors within ±2\% range. Finally, the volumes were uniformly scaled by $-5$\%, $-2.5$\%, 0\%, $+2.5$\%, and $+5$\% and were then relaxed with the FIRE optimizer \cite{bitzek2006structural} using the MACE-MP-0b3 foundation model \cite{batatia2025foundation} until the forces converged to $10^{-4}$ eV/\AA. From each relaxation trajectory, four configurations equally spaced in energy from the starting structure until 90\% of convergence were selected for subsequent DFT calculations. 
In total, 20 configurations were obtained per material. The 4 configurations originating from the (0\%) unscaled volume trajectory were assigned as validation data, while the remaining 16 configurations were used for training.
The fine-tuning data were successively added along the four relaxation trajectories, starting with only the outermost structures from each trajectory. In subsequent steps, the next inner configurations were also included, progressively expanding the training set toward the equilibrium region. Consequently, the first training data points correspond to the ±5 \% volume-scaled and distorted structures, followed by increasingly relaxed configurations that approach the energy minimum.

\subsection{Model hyperparameters}
In this study, we use the MACE-MP-0b3 foundation model, which employs a cutoff radius of 6 \AA~and two interaction layers, limiting the receptive field to 12~\AA. Models were fine-tuned using the AdamW optimizer with a ReduceLROnPlateau scheduler for up to 200 epochs. Training minimized a weighted sum of Huber losses for energies, forces, and stresses with $(w_e, w_f, w_s) = (10, 100, 1000)$ and $\delta_H = 0.01$. Transfer-learned models used $(lr, wd) = (0.01, 5\times10^{-3})$, multi-head models used $(0.01, 5\times10^{-7})$ together with a replay dataset of 100 randomly chosen structures, and Equitrain models $(0.01, 10)$. The from-scratch models are not pre-trained; instead, they are trained from scratch on the fine-tuning data using the same training parameters as the multi-head models.

\subsection{LoRA fine-tuning strategies in Equitrain}
Equitrain~\cite{equitrain_website} implements multiple fine-tuning strategies, including layer freezing and LoRA, with support for both low-rank and full-rank parameterizations.

Low-Rank Adaptation (LoRA) is a parameter-efficient fine-tuning method~\cite{hu2022lora}, in which a pre-trained weight matrix $W_0$ is kept fixed, and task adaptation is performed by learning an additive update. Rather than optimizing the full weight matrix $W$ of a neural network layer directly, LoRA parameterizes the update as
\begin{equation*}
    W = W_0 + \Delta W,
\end{equation*}
where $\Delta W$ is expressed in a factorized form $\Delta W = BA$, with $A \in \mathbb{R}^{r \times d_{\text{in}}}$ and $B \in \mathbb{R}^{d_{\text{out}} \times r}$. The rank $r$ is typically chosen such that $r \ll \min(d_{\text{in}}, d_{\text{out}})$, which substantially reduces the number of trainable parameters while preserving strong empirical performance.

When the rank $r$ is increased to match the full dimensionality of the weight matrix, the factorized update $BA$ is capable of representing any matrix in $\mathbb{R}^{d_{\text{out}} \times d_{\text{in}}}$. In this sense, full-rank LoRA has the same representational capacity as standard full fine-tuning. However, representational equivalence does not imply equivalence in optimization or inductive bias. In practical LoRA training, regularization such as weight decay is typically applied only to the additive update $\Delta W$, while the pre-trained weights $W_0$ remain frozen. The resulting optimization problem for estimating the additive weights $\Delta W$ can be written as
\begin{equation*}
    \Delta W = \argmin_{\Delta W} \; \mathcal{L}(W_0 + \Delta W) + \lambda \|\Delta W\|^2\,,
\end{equation*}
which explicitly penalizes deviation from the pre-trained solution. Here, $\mathcal{L}$ is our training objective function, and $\lambda$ is the strength of the weight decay used for regularization.

This objective differs fundamentally from standard full fine-tuning, where the model parameters are optimized directly without an explicit constraint anchoring them to the pre-trained weights. Instead, full-rank LoRA with update regularization corresponds to a form of proximal or trust-region fine-tuning, in which learning is biased toward solutions close to $W_0$. Similar regularization schemes have been studied in transfer learning under the name L2-SP~\cite{xuhong2018explicit}, where constraining parameter updates relative to pre-trained weights was shown to improve stability and generalization.

Moreover, even in the absence of an explicit rank constraint, the factorized parameterization of the update alters the optimization geometry and gradient dynamics. Recent work has emphasized that such differences in parameterization and regularization can lead to distinct training trajectories and solutions, despite identical expressive power~\cite{shuttleworth2024lora}. Consequently, full-rank LoRA with weight decay should not be viewed as equivalent to unconstrained full fine-tuning, but rather as a regularized additive adaptation method that encodes a strong inductive bias toward the pre-trained model.

Therefore, in this study, we adopt a full-rank LoRA parameterization, which removes the need to tune the rank hyperparameter while retaining the expressive capacity of unconstrained adaptation.

\subsection{Property evaluation}
\textbf{Phonons and thermodynamics:} Phonon dispersions and thermodynamic properties at 300 K were computed using \texttt{Phonopy} \cite{phonopy-phono3py-JPCM, phonopy-phono3py-JPSJ}. \\
\textbf{Elastic and equation-of-state properties.} Elastic constants and equations of state were computed using the corresponding ML models following the same protocols as for the DFT reference. In rare cases, ML-based equation-of-state fits found individual E(V) points to lie far from the physically expected energy–volume curve. These outliers were excluded from the Birch–Murnaghan equation-of-state fit.

\subsection{Structural instabilities and phase transitions}
Dynamical instabilities were identified by imaginary phonon frequencies below $-10^{-4}$ THz at commensurate $k$-points. Supercells commensurate with unstable wave vectors were displaced along the corresponding eigenmodes, and a set of static calculations was performed to find the minima. 
In case of degenerate eigenvalues, a linear combination of eigenvectors was used. The maximum eigenvalue multiplicity in the dataset was four. For one-dimensional mode subspaces, 21 static energy calculations were performed, while two-, three-, and four-dimensional subspaces were sampled on uniform grids of 21×21, 9×9×9, and 7×7×7×7 points, respectively. 
Local energy minima within each grid were identified, and the corresponding configurations were subsequently relaxed by geometric optimization. The DFT calculations were carried out with an energy convergence criterion of $10^{-7}$ eV and a cutoff energy of 400 eV. Structural optimization with the ML models was performed until the ionic forces converged below $5$ meV/\AA{}. The geometric optimization of the ML models was additionally limited to a maximum of 2,000 steps. During relaxation, symmetry lowering was forbidden. Final symmetries were determined using \texttt{spglib} \cite{spglib} with a tolerance of 0.1 \AA.

\section{Data availability}
The datasets, evaluation and fine-tuning scripts, and machine-learning models used in this work are available in a Zenodo repository at  https://doi.org/10.5281/zenodo.19354922.
\section{Acknowledgments}
J.G. was supported by ERC Grant MultiBonds (grant agreement no. 101161771; funded by the European Union. Views and opinions expressed are however those of the author(s) only and do not necessarily reflect those of the European Union or the European Research Council Executive Agency. Neither the European Union nor the granting authority can be held responsible for them.)
Furthermore, we would like to acknowledge the Gauss Centre for Supercomputing e.V. (www.gausscentre.eu) for funding this project by providing generous computing time on the GCS Supercomputer SuperMUC-NG at Leibniz Supercomputing Centre (www.lrz.de) (Project No. pn73da)
\newpage
\printbibliography
\newpage
\section{Supplementary information}
\subsection{Foundation model comparison}
\label{SI:diff_foud_models}
At the time this study was conducted, MP-0b3 was the most recent foundation model available.
However, within the MACE-family of foundation models, MACE-MP-0b3, MACE-MPA-0 \cite{batatia2025foundation}, and MACE-OMAT-0 \cite{batatia2025cross} are the
currently available state-of-the-art models. The MACE-F model is a model specifically trained for phonon calculations \cite{lee2025accelerating}. To assess the impact of foundation model choice, we evaluate the phonon band structures of the materials in our dataset using these foundation models.
As summarized in \ref{tab:SI_foundation_model_comparison},
\begin{table}[h!]
    \centering
    \caption{Phonon MAE and IQR in THz of four different foundation models.}
    \begin{tabular}{c| c| c| c} 
        \hline
        MP-0b3  & MPA-0 & MACE-OMAT-0 & MACE-F \\ \hline \hline
        0.27 (0.18) & 0.16 (0.19) & 0.10 (0.09) & 0.29 (0.23)
    \end{tabular}
    \label{tab:SI_foundation_model_comparison}
\end{table}
MACE-MPA-0 and MACE-OMAT-0 exhibit much lower error than MP-0b3, confirming that the latter is no longer the strongest available pre-trained model. A key reason for the improved performance of the newer models is the increase in training set sizes. Whereas MP-0b3 was trained on the Materials Project dataset with a total of 1.58M structures, MACE-MPA-0 was trained on an additional 10.4M structures. MACE-OMAT-0 was trained on the OMat24 dataset containing 101M structures. 
However, the Equitrain fine-tuning strategy reduces the phonon error to a median MAE (IQR) of 0.05 (0.05) THz, substantially outperforming all tested models. This demonstrates the immense potential of fine-tuning, which can also be applied to the other models in precisely the same way, probably leading to even smaller errors. The gains from material-specific fine-tuning exceed the differences among the available foundation models. The present results likely represent a conservative estimate of the performance that can be achieved with fine-tuning and especially with Equitrain.
Notably, MACE-F shows a larger error than the other models, despite being designed for phonon calculations. This further highlights the superiority of material-specific fine-tuning. 

\subsection{Element distribution in material dataset}

\subsection{Training structure size convergence}
\begin{figure}[h!]
\centering
\begin{tabular}{ccc}
  \includegraphics[width=0.28\textwidth]{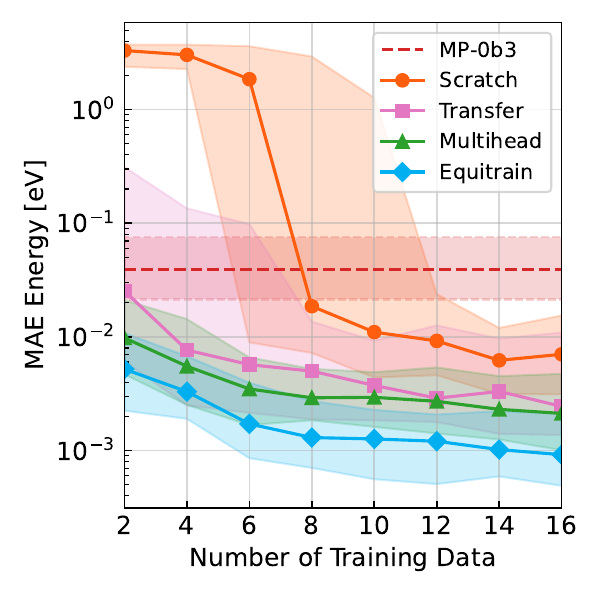} &
  \includegraphics[width=0.28\textwidth]{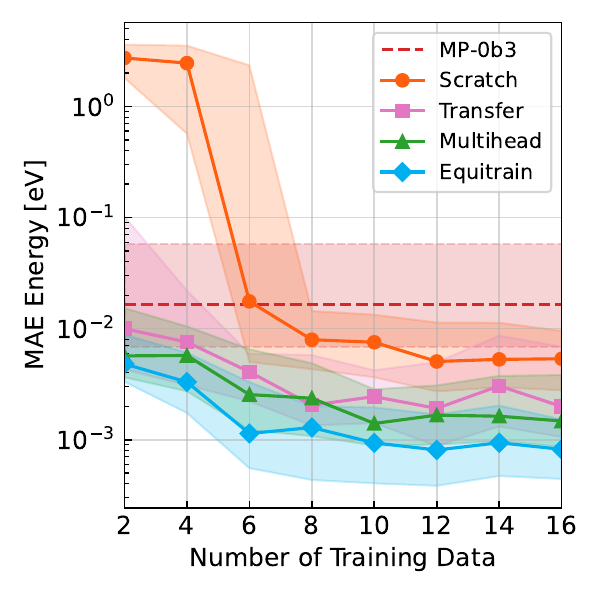} &
  \includegraphics[width=0.28\textwidth]{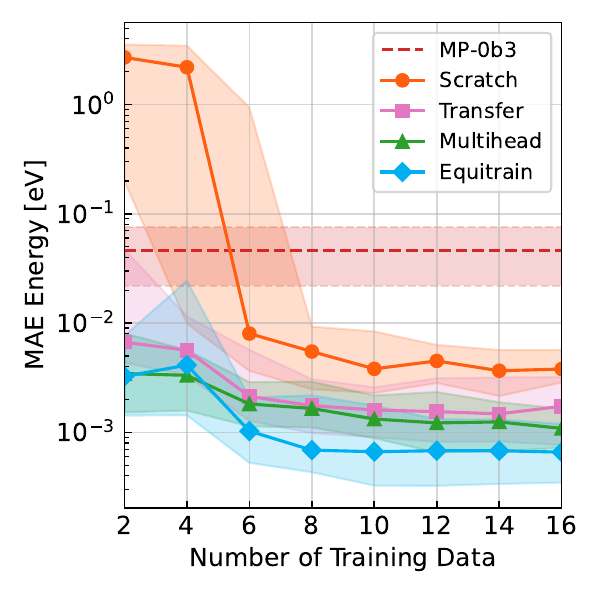} \\

  \includegraphics[width=0.28\textwidth]{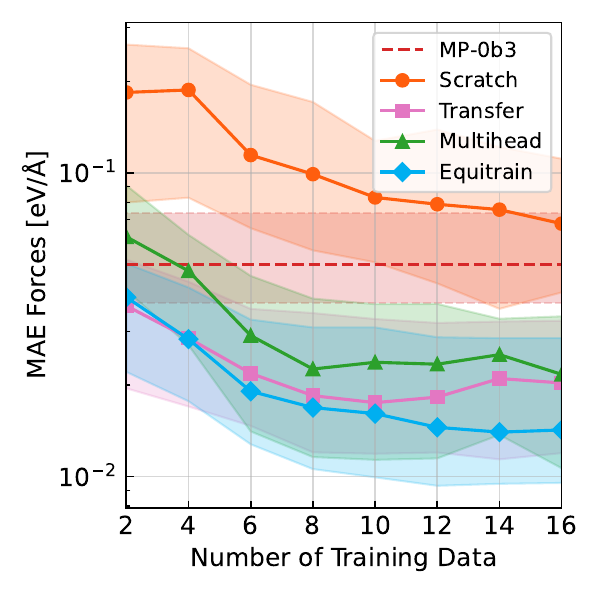} &
  \includegraphics[width=0.28\textwidth]{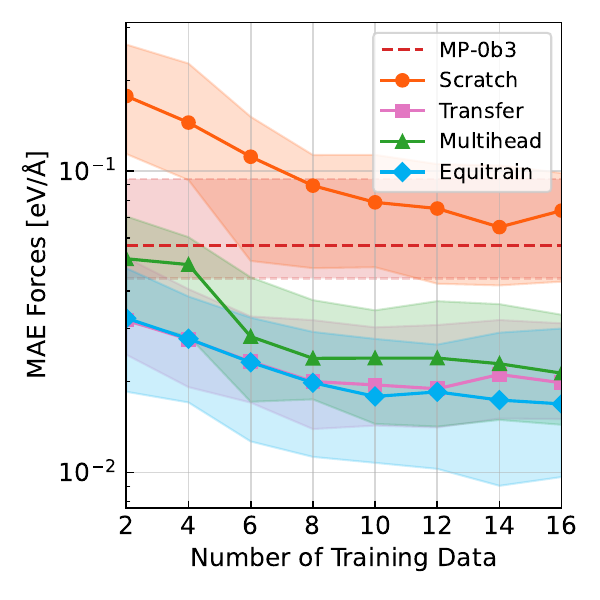} &
  \includegraphics[width=0.28\textwidth]{training_convergence/forces_super.pdf} \\

  \includegraphics[width=0.28\textwidth]{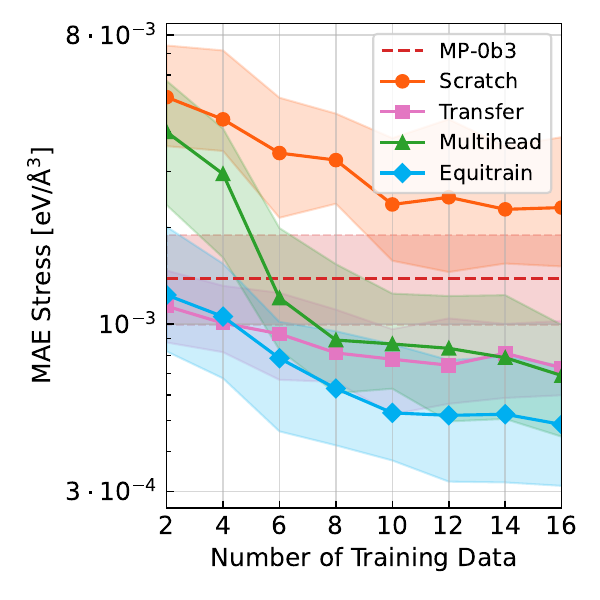} &
  \includegraphics[width=0.28\textwidth]{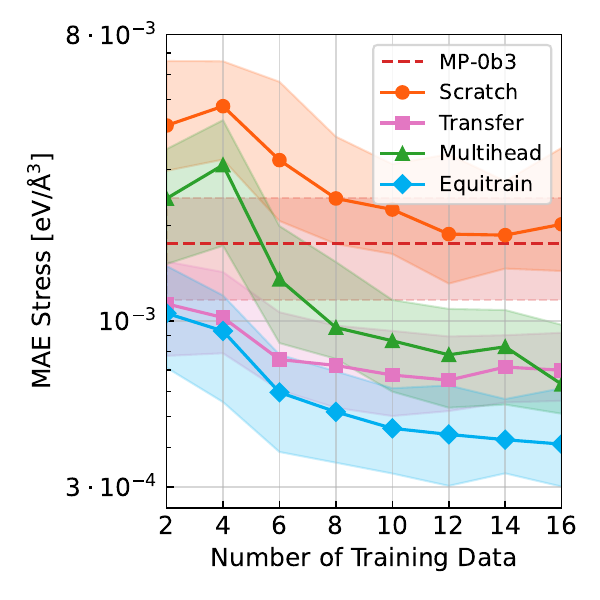} &
  \includegraphics[width=0.28\textwidth]{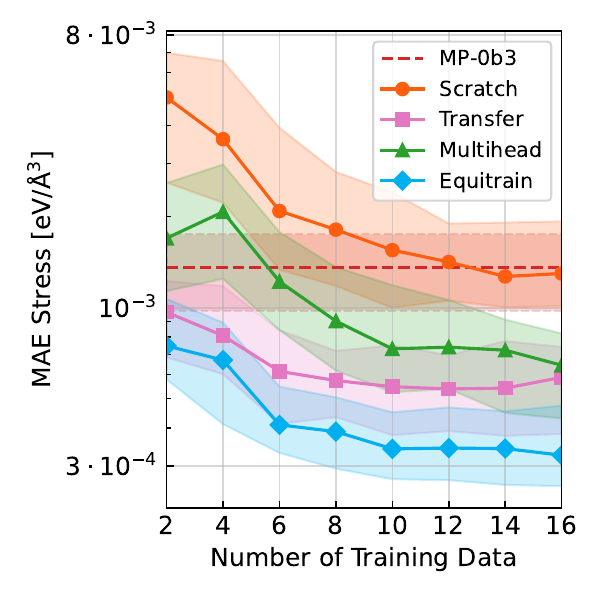} \\[0.5em]

  (a) & (b) & (c)
\end{tabular}
\caption{Energy, force, and stress median MAE and inter-quartile range on the validation sets of all 53 materials trained with a) primitive unit cells, b) small supercells, and c) large supercells. For comparison, the red dashed line shows how well the foundation MP-0b3 model performs on the validation sets.}
\label{fig:SI_training_size_convergence}
\end{figure}

Figure \ref{fig:SI_training_size_convergence} shows the Median MAE of energy, forces, and stress for training set sizes of (a) primitive unit cells, (b) small supercells, and (c) large supercells. The overall convergence trend is the same in all figures. The MAE decreases significantly with up to 10 additional fine-tuning data points. After that, the trend flattens out.

\subsection{Calculation of physical properties}
\subsection*{Thermal properties}
The heat capacity, Helmholtz free energy, and entropy are calculated directly from the phonon band structures using the implementation in the Phonopy code \cite{phonopy-phono3py-JPSJ}.
\subsubsection*{Slack thermal conductivity}
\label{SI:slack_thermal_cond}
The Slack thermal conductivity \cite{slack1973nonmetallic}
\begin{equation}
\kappa_\text{Slack} = A(\gamma)\frac{M\delta n^{1/3} \Theta_D^3}{\gamma^2T}
\end{equation}
depends on the average Grüneisen parameter $\gamma$, Here, $\gamma$ is calculated from the longitudinal and transversal sound velocities $v_L, v_T$ \cite{jia2017lattice}
\begin{equation}
    \gamma = \frac{3}{2}(\frac{1+\nu}{2-3\nu}), \qquad \nu = \frac{1-2(v_T/v_L)^2}{2-2(v_T/v_L)^2} 
\end{equation}
The sound velocities are obtained from elastic-property calculations \cite{jia2017lattice, ganose2025atomate2}.
They are calculated from the Voigt-Reuss-Hill average shear modulus $\bar{G}_{\text{VRH}}$ and bulk modulus $\bar{B}_{\text{VRH}}$
\begin{align}
    v_T &= \left(\frac{\bar{G}_{\text{VRH}}}{\rho_m}\right)^{\frac{1}{2}} \\
    v_L &= \left(\frac{\bar{B}_{\text{VRH}} + 4/3 \bar{G}_{\text{VRH}}}{\rho_m}\right)^{\frac{1}{2}}
\end{align}
where $\rho_m$ is the mass density.

\subsubsection*{Bulk modulus}
The bulk modulus was calculated by fitting the Birch-Murnaghan equation of state \cite{hebbache2004ab} to energy-volume data.
\begin{equation}
    E(\eta) = E_0 + \frac{9B_0V_0}{16}(\eta^2-1)^2(6+B_0'(\eta^2-1)-4\eta^2)
\end{equation}
with $\eta=\left(V/V_0\right)^{\frac{1}{3}}$. Here, $V$, $V_0$ are the volume and equilibrium volume, $E_0$ is the minimum energy, $B_0$ is the bulk modulus, and $B_0'$ is its first derivative with respect to pressure.

\subsubsection*{Shear modulus}
The Voigt-Reuss-Hill shear modulus $G_{\text{VRH}}$ is defined as the average of the Voigt shear modulus $G_{V}$ and the Reuss shear modulus $G_R$.
\begin{equation}
    G_{\text{VRH}} = \frac{1}{2}(G_V +G_R)
\end{equation}

\subsection{Physical properties with outliers}
\label{SI:phys_prop_with_out}
\begin{figure}[h!]
    \centering
    \includegraphics[width=0.8\linewidth]{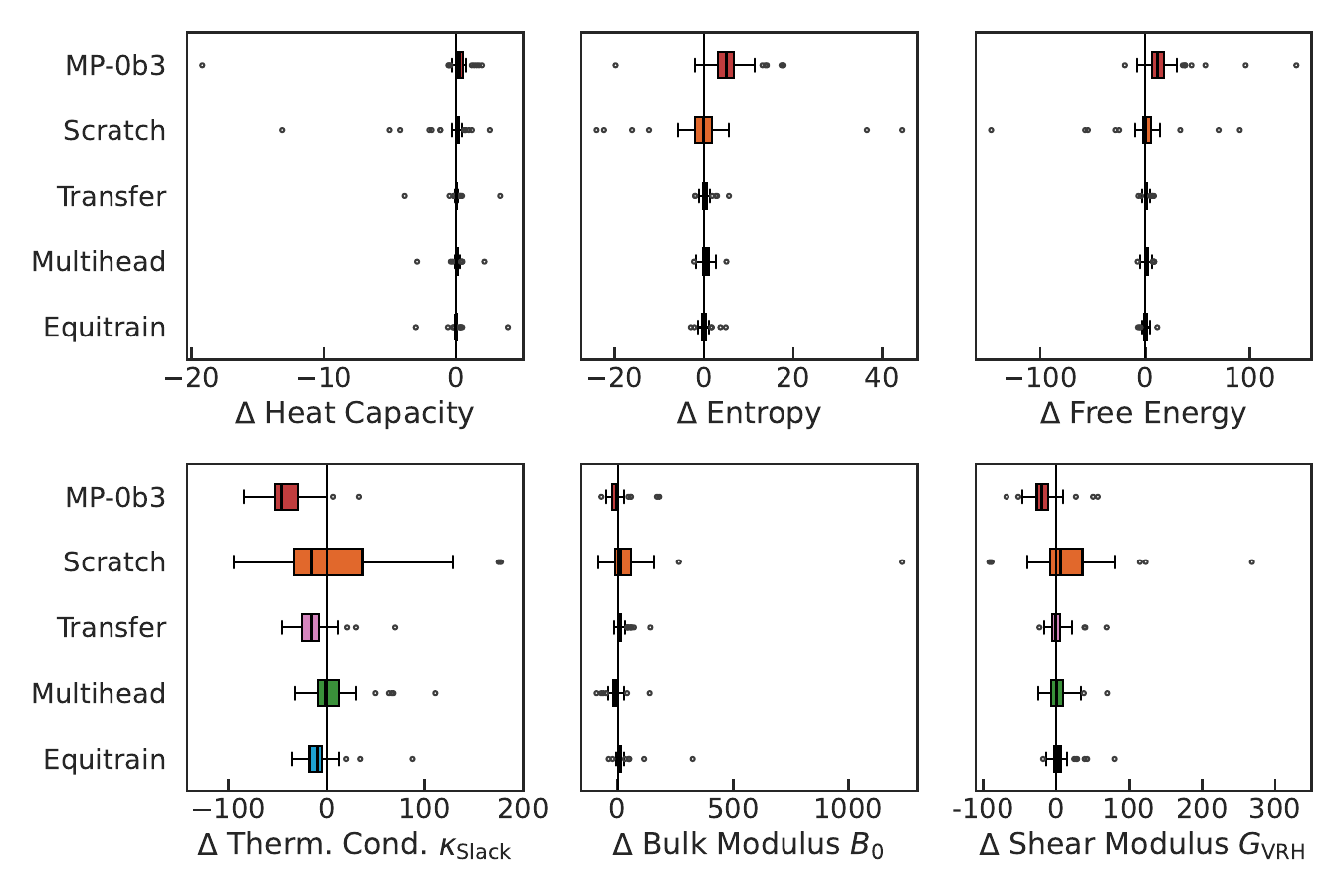}
    \caption{Deviation [\%] of thermal properties from the DFT reference results. Heat capacity, entropy, and free energy are calculated at 300 K. Boxplots including outliers are shown.}
    \label{fig:SI_boxplots_with_outliers}
\end{figure}

Figure \ref{fig:SI_boxplots_with_outliers} presents the same data as Figure \ref{fig:thermal_prop}, but with outliers included. In most cases, the outliers appear mainly in the MP-0b3 and from-scratch models, whereas all fine-tuning approaches reduce the number of outliers.

\subsection{Phase transition pathways}
\label{SI:phase-transition-pathways}
Within the finite-displacement approach, phonon band structures are obtained exactly only at $k$-points that are commensurate with the chosen supercell. The size and shape of the supercell used for single-atom displacements, therefore, directly determine which $k$-points are calculated exactly and which frequencies are obtained by Fourier interpolation.
To analyze the phase transition pathways, all imaginary modes at high-symmetry points and at minima of the phonon bands were considered. For each of these modes, the smallest supercell commensurate with the corresponding $k$-point was constructed. Only calculations with fewer than 1000 atoms per supercell were carried out. By calculating the potential energy along the corresponding imaginary modes and comparing this to ML predictions, we identify four possible scenarios.\\
\begin{enumerate}
\item DFT and ML predict identical phase transition pathways. The shapes of the energy potential and the relaxed structure obtained at the energy minimum are consistent between the two methods.
\item The DFT phonon band structure exhibits imaginary modes, but the energy potential along the corresponding mode remains purely harmonic. This behavior can be attributed to interpolation errors inherent to the finite-displacement approach.
\begin{figure}[h!]
    \subfigure[]{\includegraphics[width=0.4\textwidth]{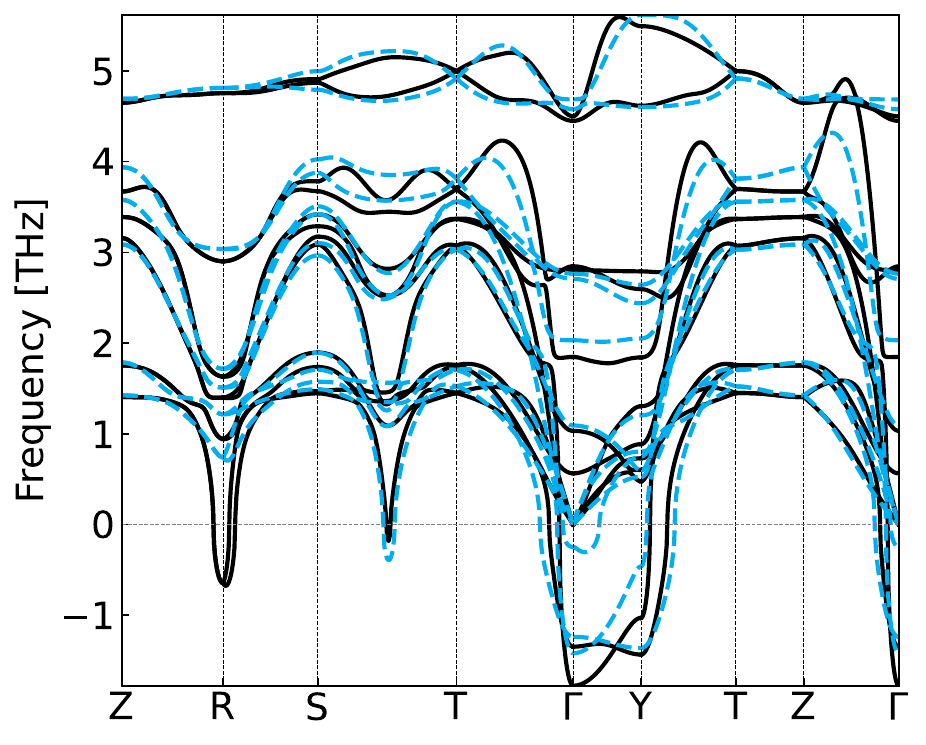}}
    \subfigure[]{\includegraphics[width=0.6\textwidth]{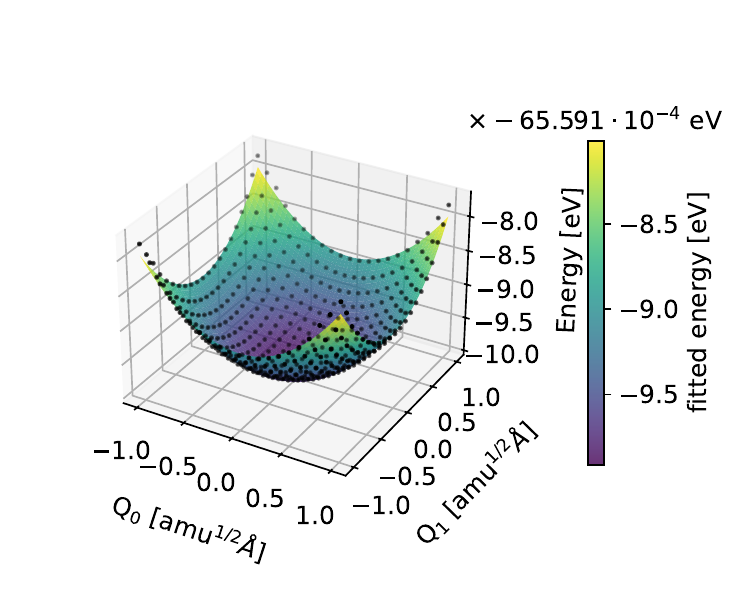}}
    \subfigure[]{\includegraphics[width=0.4\textwidth]{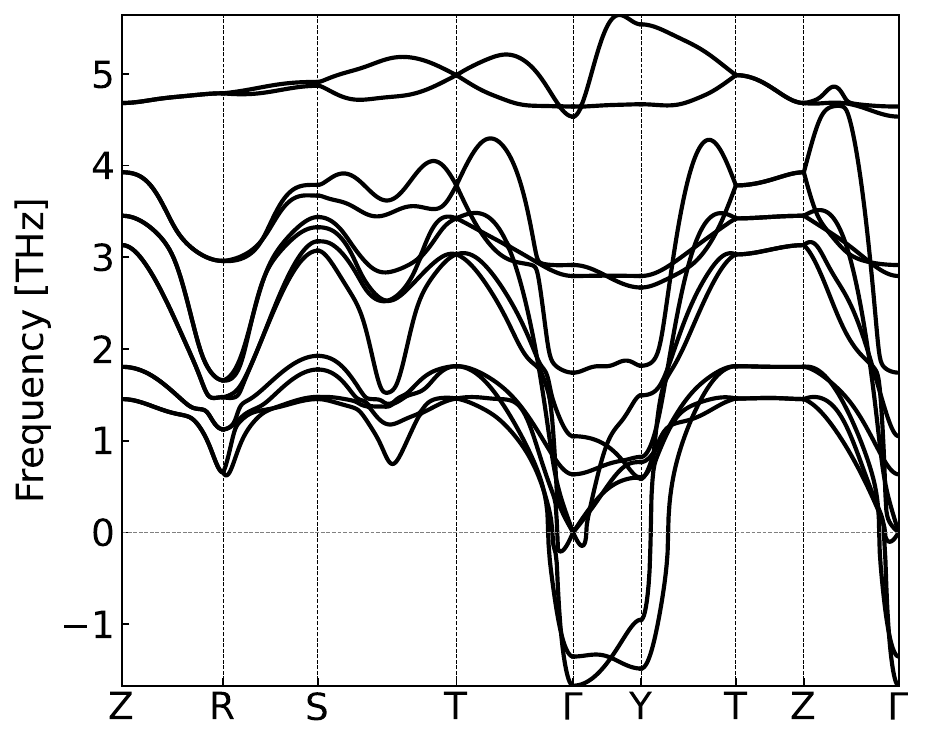}}
\caption{(a) Comparison of the full phonon band structure of $Cmcm$ SnSe between DFT (black) and the Equitrain model (blue). (b) shows the harmonic energy potential for displacements along the imaginary $R$-mode. (c) DFT calculation with a different supercell matrix.}
\label{fig:SI_mp-2168}
\end{figure}

An example is shown in Figure \ref{fig:SI_mp-2168} (a). The $k$-point $R=(0,0.5,0.5)$ is not commensurate with the chosen supercell matrix  
\[S_1 = \begin{pmatrix}
5 & -2 & 0 \\
5 &  2 & 0 \\
0 &  0 & 5
\end{pmatrix}\]
The corresponding energy potential along the degenerate modes is shown in Figure \ref{fig:SI_mp-2168} (b). The black dots represent the calculated DFT energies, while the colored surface shows the fitted harmonic energy function
\begin{equation}
E(Q_0,Q_1)= -65.59199 \text{ eV} + 9.8\cdot10^{-5} \frac{\text{eV}}{\text{amu\AA$^2$}}Q_0^2 + 9.8\cdot10^{-5}\frac{\text{eV}}{\text{amu\AA$^2$}}Q_1^2 
\end{equation}
with $R^2=0.99$.
By choosing a different supercell matrix   
\[S_2 = \begin{pmatrix}
5 & -2 & 0 \\
5 &  2 & 0 \\
0 &  0 & 4
\end{pmatrix}\]
the single-atom displaced supercells become commensurate with $R$. The resulting phonon band structure, shown in Figure \ref{fig:SI_mp-2168} (c), confirms that the imaginary mode in (a) originated from Fourier interpolation errors. \\
The $k$-point (0.02,0.02,0) between $S$ and $T$ would require a large supercell of size $50\times50\times1$ to calculate the potential energy, which is computationally prohibitive and therefore not tested.
\item As in scenario 2, DFT exhibits imaginary modes due to interpolation errors. However, the ML model captures an anharmonic potential and relaxes to the same primitive unit cell with same space group, differing only by slightly adjusted lattice parameters. It should be noted here that the energy potential is not correctly represented by the ML model, as a minimally displaced structure is found along the anharmonic direction. 
\item Both DFT and ML models exhibit imaginary modes, but the ML model follows an incorrect transition pathway and relaxes into a different space group than obtained from DFT. An example for this is the transition pathway of K$_3$Sb in the Multihead Model, which was discussed in the main part in \ref{results_imaginary_mode_pathways}.
\end{enumerate}
For each material, the set of final phases, $S_{ML}$, is compared to the set of DFT reference phases, $S_{DFT}$, as follows:
First, the initial phase is removed from both sets of final phases, because it does not represent a new displacive phase and typically corresponds to either scenario 2 or scenario 3.
Thus, for an arbitrary phase transition pathway $A \rightarrow B \rightarrow C$, the initial phase $A$ is excluded from both sets $S_{ML}$ and $S_{DFT}$.  \\
\begin{align}
    S_{ML} &\leftarrow S_{ML}\setminus\{A\} \\
    S_{DFT} &\leftarrow S_{DFT}\setminus\{A\}
\end{align}
The remaining phases are classified as follows:  
\begin{itemize}
    \item $P\in S_{ML}$ is labeled as true positive, if $P\in S_{DFT}$.
    \item $P\in S_{ML}$ is labeled as false positive if $P\notin S_{DFT}$.
    \item $P\in S_{DFT}$ is labeled as false negative, if $P\notin S_{ML}$.
\end{itemize}
The following tables list all imaginary modes and their corresponding phase-transition pathways. "imag" denotes the total number of imaginary modes found at band minima or high-symmetry $k$-points. Due to the restriction of a maximum of 1000 atoms per commensurate supercell, only those phase transitions that could be explicitly calculated are shown in the table. Empty entries indicate the absence of imaginary modes. Entries in the DFT Table \ref{tab:SI_dft_table} labeled as "single minimum" indicate the presence of a single (harmonic) minimum as described in scenario 2.
"Initial sym." and "\#Atoms" denote the space group symmetry and number of atoms in the primitive unit cell of the initial structure, respectively. \\
"displ. Sym." and "final Sym." refer to the crystal symmetry of the displaced structure at the energy minimum and of the final structure after full geometric optimization, respectively. 
"$\Delta E$/Atom" denotes the energy difference between the initial and final structure.

\begin{table*}[ht]
\centering
\caption{DFT}
\small
\resizebox{\textwidth}{!}{
\begin{tabular}{lllllllll}
\hline
material & imag. & $k$-point & initial sym. & \#atoms & displ. sym. & final sym. & \#atoms & $\Delta E$/atom [eV] \\
\hline

SnSe (mp-691) & 1 & 0.15/0/0 & Pnma & 2 & Pnma & Pnma & 2 & 0.00011886 \\ \hline

SnTe (mp-1883) & 2 & 0/0/0 & Fm$\overline{3}$m & 2 & Cm & Cm & 2 & 0.003926925 \\
 & & 0.5/0/0.5 & Fm$\overline{3}$m &  & P4/nmm & P4/nmm & 4 & 0.000893576 \\ \hline

Sb$_2$Se$_3$ (mp-2160) & 4 & 0/0/0 & Pnma & 20 & P2$_1$2$_1$2$_1$ & P2$_1$2$_1$2$_1$ & 20 & 0.000234258 \\
 & & 0/0/0.35 & & & \multicolumn{2}{c}{single minimum}  \\
 & & 0.5/0/0 & Pnma &  & P2$_1$/c & P2$_1$/c & 40 & 0.000332583 \\ \hline

SnSe (mp-2168) & 7 & -0.5/0.5/0 & Cmcm & 4 & Pnma & Pnma & 8 & 0.006528244 \\
 & & & & & Pnma & Pnma & 8 & 0.000847502 \\
 & & 0/0/0 &  &  & Cmc2$_1$ & Cmc2$_1$ & 4 & 0.006847633 \\
 & & & &  & Ama2 & Ama2 & 4 & 0.002075974 \\
 & & 0/0.5/0.5 & & & \multicolumn{2}{c}{single minimum} \\
 & & 0.25/0.25/0.5 & & & Cmcm &  Cmcm & 4 & 0.009638485 \\ \hline

GeTe (mp-2612) & 5 & 0/0/0 & Fm$\overline{3}$m & 2 & R3m & R3m & 2 & 0.01955798 \\
 & & 0.5/0/0.5 & Fm$\overline{3}$m &  & Pmmn & Pmmn & 4 & 0.003247811 \\
 & & & Fm$\overline{3}$m &  & P4/nmm & P4/nmm & 4 & 0.000450387 \\
 & & 0.5/0.25/0.75 & & & \multicolumn{2}{c}{single minimum} \\
 & & 0.375/0.375/0.75 & & & \multicolumn{2}{c}{single minimum} \\
 & & 0.625/0.25/0.625 & & & Fm$\overline{3}$m & Fm$\overline{3}$m & 2 & 0.00001452 \\ \hline

K$_3$Sb (mp-14017) & 1 & 0.333/0.333/0.333 & P6$_3$/mmc & 8 & P6$_3$cm & P6$_3$cm & 24 & 0.00090241 \\ \hline

As$_2$PbS$_4$ (mp-19941) & 12 & 0/0.5/0 & Pnma & 28 & P2$_1$/c & P2$_1$/c & 56 & 0.020323653 \\
 & & &         & & P2$_1$/c & P2$_1$/c & 56 & 0.013101457 \\
 & & 0/0.5/0.5 & & & P2$_1$/c & P2$_1$/c & 56 &  0.015168392 \\
 & &           & & & P2$_1$/c & P2$_1$/c & 56 & 0.015913025 \\
 & & 0.5/0/0 & & & Pnma & Pna2$_1$ & 28 & 0.010950972 \\
 & & 0.5/0/0.5 & & & P2$_1$/c & P2$_1$/c & 56 & 0.011384632\\
 & & 0.5/0.5/0 & & & P$\overline{1}$ & P$\overline{1}$ & 56 & 0.019336829 \\
 & & 0.5/0.5/0.5 & & & Cc & Cc & 56 & 0.015905725 \\ \hline

Bi$_2$S$_3$ (mp-22856) & 6 & 0/0/0 & Pnma & 20 & Pnma & Pnma & 20 & 0.00128557 \\
 & & 0/0.05/0 & & & \multicolumn{2}{c}{single minimum} \\
 & & 0.5/0/0 & &  & P2$_1$/c & P2$_1$/c & 40 & 0.000950493 \\ \hline

Na$_2$TlSb (mp-866132) & 3 & 0.5/0/0.5 & Fm$\overline{3}$m & 4 & Pmmn & Pmmn & 8 & 0.007805488 \\ \hline
\end{tabular}
}
\label{tab:SI_dft_table}
\end{table*}

\begin{table*}[h!]
\centering
\caption{MP-0b3}
\small
\resizebox{\textwidth}{!}{
\begin{tabular}{lllllllll}
\hline
material & imag. & $k$-point & initial sym. & \#atoms & displ. sym. & final sym. & \#atoms & $\Delta E$/atom [eV] \\
\hline

SnSe (mp-691) & 0 & & & & & & &  \\ \hline

SnTe (mp-1883) & 1 & 0/0/0 & Fm$\overline{3}$m & 2 & P1 & R3m & 2 & 0.00356861 \\ \hline

Sb$_2$Se$_3$ (mp-2160) & 3 & 0/0/0 & Pnma & 20 & P2$_1$2$_1$2$_1$ & Pnma & 40 & 0.008583621 \\ 
 & & 0/0.5/0 & & & P2$_1$/c  & P2$_1$/c & 40 & 0.00858354 \\ \hline

SnSe (mp-2168) & 12 & -0.5/0.5/0 & Cmcm & 4 & Pnma & Pnma & 8 & 0.001517438 \\
 & & 0/0/0 &  &  & Ama2 & Ama2 & 4 & 0.002312192 \\
 & & &  & & Cmc2$_1$ & Cmc2$_1$ & 4 & 0.000162019 \\
 & & 0/0/0.025 &  & & Ama2 & Ama2 & 160 & 0.001510682 \\
 & & &  & & Cmcm & Cmcm & 4 & 0.0000000079 \\
 & & &  & & Ama2 & Ama2 & 160 & 0.0000526411 \\
 & & 0/0.005/0 &  &  & Cmcm & Cmcm & 4 & 0.0000000079 \\
 & & &  &  & Cmcm & Cmcm & 4 & 0.0000000079 \\
 & & 0/0.5/0.5 &  & & P1 & Cm & 8 & 0.000619386 \\ \hline

GeTe (mp-2612) & 5 & 0/0/0 & Fm$\overline{3}$m & 2 & Cm & R3m & 2 & 0.019445665 \\
 & & 0.5/0/0.5 & &  & P4/nmm & P4/nmm & 4 & 0.00430617 \\
 & & &  &  & P2$_1$/m & Pmmn & 4 & 0.00340047 \\
 & & 0.5/0.25/0.75 & &  & C2/m & C2/m & 8 & 0.000002035 \\
 & & 0.375/0.375/0.75 &  &  & Pmma & Pmma & 16 & 0.002372796 \\
 & & 0.625/0.25/0.625 & &  & Pmma & Pmma & 16 & 0.002372796 \\ \hline

K$_3$Sb (mp-14017) & 0 & & & & & & & \\ \hline

As$_2$PbS$_4$ (mp-19941) & 22 & 0/0/0 & Pnma & 28 & P2$_1$/c & P2$_1$/c & 28 & 0.006660053 \\
 & & &  &  & P2$_1$/c & P2$_1$/c & 28 & 0.013471486 \\
 & & 0/0/0.5 &  &  & Pca2$_1$ & Pca2$_1$ & 56 & 0.009556708 \\
 & & 0/0.5/0 &  &  & P2$_1$/c & P2$_1$/c & 56 & 0.009728751 \\
 & & &  &  & P2$_1$/c & P2$_1$/c & 56 & 0.010450992 \\
 & & 0/0.5/0.5 &  &  & P2$_1$/c & P2$_1$/c & 56 & 0.010735403 \\
 & & &  &  & P2$_1$/c & P2$_1$/c & 56 & 0.011743375 \\
 & & 0.5/0/0 &  &  & P2$_1$/c & P2$_1$/c & 56 & 0.011743375 \\
 & & &  &  & Pna2$_1$ & Pna2$_1$ & 56 & 0.005817168 \\
 & & 0.5/0/0.5 &  &  & P2$_1$/c & P2$_1$/c & 56 & 0.008533517 \\
 & & &  &  & P2$_1$/c & P2$_1$/c & 56 & 0.006087639 \\
 & & 0.5/0.4/0 &  &  & P2$_1$ & P2$_1$ & 280 & 0.011595486 \\
 & & 0.5/0.5/0 &  &  & P1 & P2$_1$ & 56 & 0.012269718 \\
 & & 0.5/0.5/0.5 &  &  & P1 & P$\overline{1}$ & 56 & 0.012249352 \\ \hline

Bi$_2$S$_3$ (mp-22856) & 2 & 0/0/0 & Pnma & 20 & P2$_1$2$_1$2$_1$ & Pnma & 20 & 0.00701234 \\
 & & 0.5/0/0 &  &  & Pc & P2$_1$/c & 40 & 0.00701227 \\ \hline

Na$_2$TlSb (mp-866132) & 3 & 0.5/0/0.5 & Fm$\overline{3}$m & 4&Fm$\overline{3}$m & Pmmn & 8 & 0.014958574 \\
 & & 0.5/0.5/0.5 &  & & C2/m & C2/m & 8 & 0.003494006 \\
 & & 0.5/0.25/0.75 &  &  & C2 & I$\overline{4}$2m & 16 & 0.006546509 \\
 & & & &  & I4/mmm & I4/mmm & 16 & 0.002032898 \\ \hline
\end{tabular}
}
\end{table*}

\begin{table*}[h!]
\centering
\caption{From-Scratch}
\small
\resizebox{\textwidth}{!}{
\begin{tabular}{lllllllll}
\hline
material & imag. & $k$-point & initial sym. & \#atoms & displ. sym. & final sym. & \#atoms & $\Delta E$/atom [eV] \\
\hline

SnSe (mp-691) & 0 & & & & & & &  \\ \hline

SnTe (mp-1883) & 0 & & & & & & &  \\ \hline

Sb$_2$Se$_3$ (mp-2160) & 2 & 0/.025/0 & Pnma & 20 & Pmn2$_1$ & Pmn2$_1$ & 800 & 0.001253666 \\ \hline

SnSe (mp-2168) & 3 & -0.5/0.5/0 & Cmcm & 4 & Pnma & Fm$\overline{3}$m & 2 & 0.015364941 \\
 & & 0/0.04/0 &  &  & Pm & P3m1 & 50 & 0.014858382 \\ \hline

GeTe (mp-2612) & 5 & 0/0/0 & Fm$\overline{3}$m & 2 & P1 & R3m & 2 & 0.005260166 \\
 & & 0.5/0/0.5 &  &  & Pmmn & Pmmn & 4 & 0.006488827 \\
 & & 0.375/0.375/0.75 &  &  & Pmma & Pmma & 16 & 0.003595268 \\
 & & 0.625/0.25/0.625 &  &  & Pmma & Pmma & 16 & 0.003595268 \\ \hline

K$_3$Sb (mp-14017) & 0 & & & & & & & \\ \hline

As$_2$PbS$_4$ (mp-19941) & 0 & & & & & & & \\ \hline

Bi$_2$S$_3$ (mp-22856) & 3 & 0/0/0 & Pnma & 20 & P2$_1$2$_1$2$_1$ & P2$_1$2$_1$2$_1$ & 20 & 0.001173041 \\
 & & 0/0.075/0 &  &  & Pmc2$_1$ & Pmc2$_1$ & 800 & 0.000315138 \\
 & & 0.5/0/0 &  &  & P2$_1$/c & P2$_1$/c & 20 & 0.001213447 \\ \hline

Na$_2$TlSb (mp-866132) & 3 & 0.5/0/0.5 & Fm$\overline{3}$m & 4 & P2$_1$/m & Pmmn & 8 & 0.006111897 \\ \hline

\end{tabular}
}
\end{table*}

\begin{table*}[h!]
\centering
\caption{Equitrain}
\small
\resizebox{\textwidth}{!}{
\begin{tabular}{llllllllll}
\hline
material & imag. & $k$-point & initial sym. & \#atoms & displ. sym. & final sym. & \#atoms & $\Delta E$/atom [eV] \\
\hline

SnSe (mp-691)   & 0  &  &  &  &  & & &  \\ \hline

SnTe (mp-1883)  & 1  & 0/0/0 & Fm$\overline{3}$m & 2 & I4mm & I4mm &  2& $0.000063844$ \\ \hline

Sb2Se3 (mp-2160)  & 4  & 0/0/0 & Pnma &  20 & P2$_1$2$_1$2$_1$ & Pnma & 20 & 0.01031492 \\
         &  & 0.5/0/0 &  &  & Pc & Pnma & 20 & 0.010314876 \\ \hline

SnSe (mp-2168)  & 3 & -0.5/0.5/0 & Cmcm & 4 & Pnma & Pnma & 8 & 0.014018347 \\
         &   &            &  &  & Pnma & Pnma & 8 & 0.000127048 \\
         &     & 0/0/0       &  &  & Cmc2$_1$ & Cmc2$_1$ & 4 & 0.006666645 \\
         &     &             &  &  & Ama2 & Ama2 & 4 & 0.003328276 \\
         &    &             &  &  & P2$_1$/m & P2$_1$/m & 4 & 0.003164842 \\ \hline

GeTe (mp-2612)  & 2 & 0/0/0 & Fm$\overline{3}$m & 2 & P1 & R3m & 2 & 0.01579701 \\
         &   & 0.5/0/0.5 & & & P2$_1$/m & Pmmn & 4 & 0.000536886 \\ \hline

K$_3$Sb (mp-14017) & 1 & 0.333/0.333/0.333 & P6$_3$/mmc & 8 & P6$_3$cm & P6$_3$cm & 24 & 0.001109341 \\ \hline

As$_2$PbS$_4$ (mp-19941) & 14  & 0/0.5/0 & Pnma & 28 & P2$_1$/c & P2$_1$/c & 56 & 0.018658018 \\
         &      &         &  &  & P2$_1$/c & P2$_1$/c & 56 & 0.013633521 \\
         &      & 0/0.5/0.5 &  &  & P2$_1$/c & P2$_1$/c & 56 & 0.014360129 \\
         &      &           &  &  & P2$_1$/c & P2$_1$/c & 56 & 0.014077855 \\
         &      & 0.5/0.5/0 &  &  & P1 & P2$_1$ & 56 & 0.017369918 \\
         &      & 0.5/0.5/0.5 &  &  & P1 & Cc & 56 & 0.018978289 \\ \hline

Bi$_2$S$_3$ (mp-22856) & 4 & 0/0/0 & Pnma & 20 & P2$_1$2$_1$2$_1$ & P2$_1$2$_1$2$_1$ & 20 & 0.008654072 \\
         &     & 0/0.05/0 &  &  & Pmc2$_1$ & Pmc2$_1$ & 400 & 0.000915012 \\
         &     & 0.5/0/0 &  &  & Pc & P2$_1$/c & 40 & 0.008653308 \\ \hline

Na$_2$TlSb (mp-866132) & 4  & 0.5/0/0.5 & Fm$\overline{3}$m & 4 & P2$_1$/m & Pmmn & 8 & 0.007704081 \\ \hline

\hline
\end{tabular}
}
\end{table*}

\begin{table*}[h!]
\centering
\caption{Multihead}
\small
\resizebox{\textwidth}{!}{
\begin{tabular}{lllllllll}
\hline
material & imag. & $k$-point & initial sym. & \#atoms & displ. sym. & final sym. & \#atoms & $\Delta E$/atom [eV] \\
\hline

SnSe (mp-691) & 0 & & & & & & & \\ \hline

SnTe (mp-1883) & 1 & 0/0/0 & Fm$\overline{3}$m & 2 & Cm & Imm2 & 2 & 0.000178059 \\ \hline

Sb$_2$Se$_3$ (mp-2160) & 4 & 0/0/0 & Pnma & 20 & Pmc2$_1$ & Pmc2$_1$ & 20 & 0.006379553 \\
 & & 0/0/0.05 &  & & Pm & Pmc2$_1$ & 400 & 0.00611383 \\
 & & &  & & Pm & Pm & 400 & 0.00282019 \\
 & & 0.5/0/0 &  & & Pm & Pm & 40 & 0.006378815 \\ \hline

SnSe (mp-2168) & 5 & -0.5/0.5/0 & Cmcm & 4 & Pnma & Pnma & 8 & 0.031494809 \\
 & & &  & & Pnnm & Pnnm & 8 & 0.024060346 \\
 & & 0/0/0 &  & & Cmc2$_1$ & Cmc2$_1$ & 4 & 0.012965835 \\
 & & &  & & Ama2 & Ama2 & 4 & 0.00548828 \\ \hline

GeTe (mp-2612) & 2 & 0/0/0 & Fm$\overline{3}$m & 2 & Cm & R3m & 2 & 0.015541084 \\
 & & 0.5/0/.5 &  & & P2$_1$/m & Pmmn & 4 & 0.002914793 \\
 & & &  & & P4/nmm & P4/nmm & 4 & 0.000237267 \\ \hline

K$_3$Sb (mp-14017) & 1 & 0.333/0.333/0.333 & P6$_3$/mmc & 8 & R-3 & R-3 & 24 & 0.000121777 \\ \hline

As$_2$PbS$_4$ (mp-19941) & 28 & 0.5/0.5/0 & Pnma & & P1 & P$\overline{1}$ & 56 & 0.024973003 \\
 & & 0.5/0.5/0.5 &  & & P1 & P$\overline{2}$ & 56 & 0.023414041 \\ \hline

Bi$_2$S$_3$ (mp-22856) & 4 & 0/0/0 & Pnma & 20 & P2$_1$2$_1$2$_1$ & Pnma & 20 & 0.011353954 \\
 & & 0.5/0/0 & & & Pc & Pnma & 20 & 4.388049686 \\ \hline

Na$_2$TlSb (mp-866132) & 4 & & & & & & & \\ \hline

\end{tabular}
}
\end{table*}

\begin{table*}[h!]
\centering
\caption{Transfer}
\small
\resizebox{\textwidth}{!}{
\begin{tabular}{lllllllll}
\hline
material & imag. & $k$-point & initial sym. & \#atoms & displ. sym. & final sym. & \#atoms & $\Delta E$/atom [eV] \\
\hline

SnSe (mp-691) & 0 & & & & & & &  \\ \hline

SnTe (mp-1883) & 0 & & & & & & &  \\ \hline

Sb$_2$Se$_3$ (mp-2160) & 4 & 0/0/0 & Pnma & 20 & P2$_1$2$_1$2$_1$ & P2$_1$2$_1$2$_1$ & 2 & 0.002272407 \\
 & & 0.5/0/0 &  &  & Pc & P2$_1$/c & 40 & 0.002273069 \\
 & & 0.375/0/0 &  &  & Pc & Pc & 160 & 0.001846368 \\
 & & &  &  & Pc & Pc & 160 & 0.001826809 \\ \hline

SnSe (mp-2168) & 5 & 0/0/0 & Cmcm & 4 & Ama2 & Ama2 & 4 & 0.004294459 \\
 & & &  &  & Cmc2$_1$ & Cmc2$_1$ & 4 & 0.005555749 \\
 & & -0.5/0.5/0 &  &  & Pnma & Pnma & 8 & 0.010656904 \\
 & & 0/0.025/0 &  &  & Pc & P2$_1$/c & 160 & 0.003972777 \\
 & & &  &  & Pm & Pm & 160 & 0.002505208 \\
 & & &  &  & Pm & Pm & 160 & 0.002503663 \\ \hline

GeTe (mp-2612) & 2 & 0/0/0 & Fm$\overline{3}$m & 2 & P1 & R3m & 2 & 0.014341852 \\
 & & 0.5/0/0.5 &  &  & Pmmn & Pmmn & 4 & 0.000859241 \\ \hline

K$_3$Sb (mp-14017) & 0 & & & & & & &  \\ \hline

As$_2$PbS$_4$ (mp-19941) & 12 & 0/0.5/0 & Pnma & 28 & P2$_1$/c & P2$_1$/c & 56 & 0.016478829 \\
 & & &  & & P2$_1$/c & P2$_1$/c & 56 & 0.013351529 \\
 & & 0.5/0.5/0 &  & & P1 & P2$_1$ & 56 & 0.015471764 \\
 & & 0.5/0.5/0.5 &  & & P1 & P$\overline{1}$ & 56 & 0.015489667 \\
 & & 0/0.5/0.5 &  &  & P2$_1$/c & P2$_1$/c & 56 & 0.01542228 \\
 & & &  & & P2$_1$/c & P2$_1$/c & 56 & 0.013614605 \\ \hline

Bi$_2$S$_3$ (mp-22856) & 4 & 0/0/0 & Pnma & 20 & P2$_1$2$_1$2$_1$ & P2$_1$2$_1$2$_1$ & 20 & 0.004766275 \\
 & & 0.5/0/0 &  &  & Pc & P2$_1$/c & 40 & 0.004766028 \\ \hline

Na$_2$TlSb (mp-866132) & & 0.5/0.25/0.75 & Fm$\overline{3}$m & 4 & C2 & Fddd & 16 & 0.000099165 \\ \hline

\end{tabular}
}
\end{table*}

\clearpage
\subsection{Freezing-layer approach}
\label{SI:freezing-layer}
\begin{figure}[h!]
\centering
\begin{tabular}{ccc}
  \includegraphics[width=0.28\textwidth]{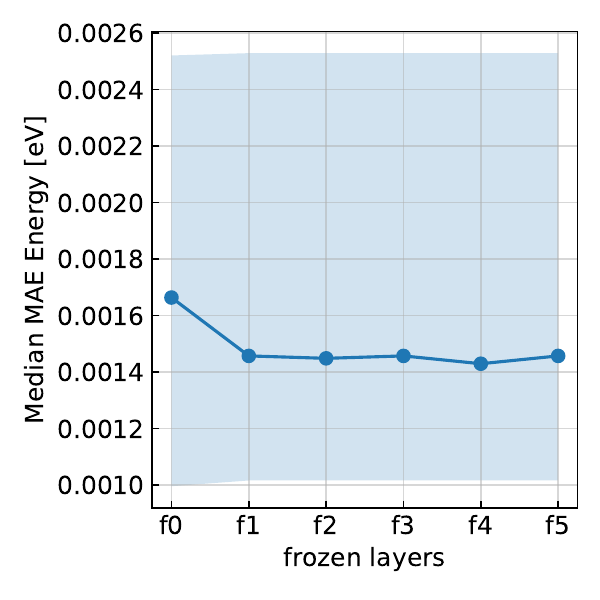} &
  \includegraphics[width=0.28\textwidth]{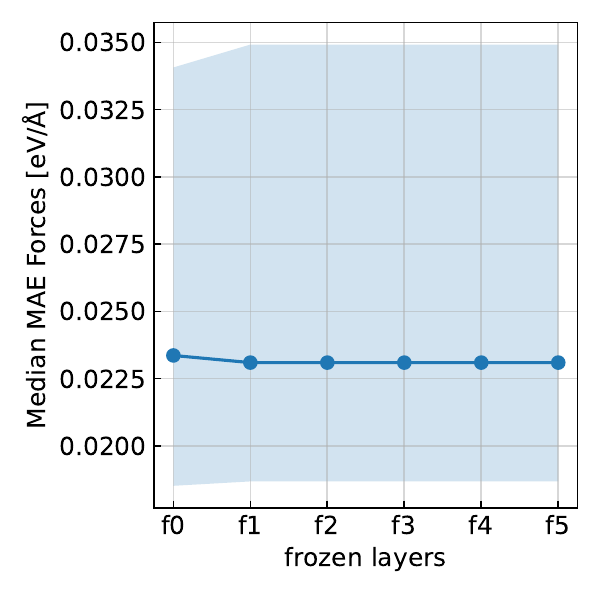} &
  \includegraphics[width=0.28\textwidth]{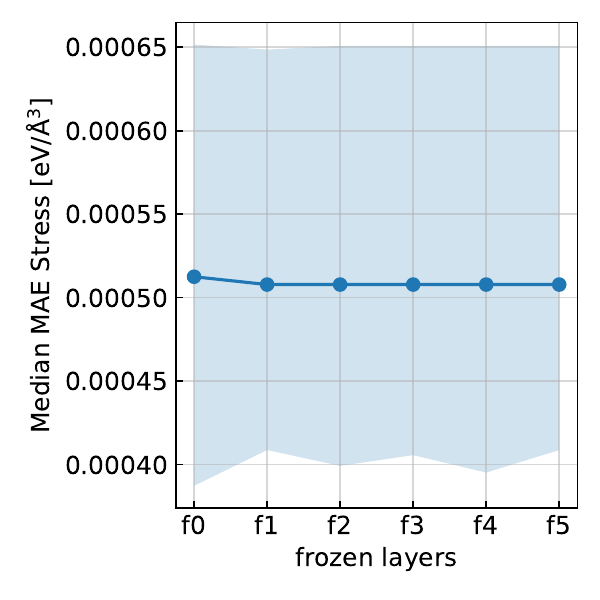} \\
\end{tabular}
\caption{Energy, force, and stress median MAE and IQR on the validation sets of all 53 materials trained with large supercells.}
\label{fig:SI_frozen_layer_convergence}
\end{figure}

The freezing-layer approach \cite{radova2025fine} is particularly useful for large datasets, as it can significantly accelerate fine-tuning with minimal loss of accuracy. In the present case, however, the fine-tuning datasets are small, such that training speed is not a limiting factor. In Figure \ref{fig:SI_frozen_layer_convergence}, the median MAE of energy, forces, and stress varies only slightly with the number of frozen layers. Here, f\{$n$\} indicates that the first $n$ layers are frozen. The largest effect is observed when freezing only the node embedding layer (f1). Freezing additional layers up to the atomic energies layer (f5) does not lead to any further improvement. \\
Since freezing the first layer shows a small but noticeable effect, we additionally computed phonons for these models.
The median phonon MAE and IQR lie at 0.08 (0.07) THz, which is slightly larger than the comparable Transfer learning approach with 0.06 (0.05) THz. For this reason, we did not pursue this approach further.

\end{document}